\documentclass[12pt]{article}
\usepackage[utf8]{inputenc}
\usepackage{amsmath, natbib}
\usepackage{graphicx}
\usepackage{amsfonts}
\usepackage{longtable}
\usepackage{colortbl} 
\usepackage{lipsum, fullpage}
\usepackage{array,multirow}
\usepackage[dvipsnames]{xcolor}
\usepackage{setspace}
\usepackage{authblk, fullpage, color, bm, soul}

\usepackage[top=2.54cm, bottom=2.54cm, left=2.2cm, right=2.2cm]{geometry}

\newcommand{\eps}{\epsilon}
\newcommand{\beps}{\bm \epsilon}

\newcommand{\iidsim}{{\overset{\mathrm{i.i.d.}}{\sim}}}

\newcommand{\balpha}{{\boldsymbol{\alpha}}}
\newcommand{\tTheta}{{\tilde{\Theta}}}

\newcommand{\bY}{{\bm{Y}}}

\newcommand{\bX}{{\bm{X}}}
\newcommand{\bLambda}{{\bm{\Lambda}}}

\newcommand{\bB}{{\bm{B}}}

\newcommand{\bR}{{\bm{R}}}

\newcommand{\bZ}{{\bm{Z}}}

\newcommand{\bt}{{\bm{t}}}

\newcommand{\bz}{{\bm{z}}}

\newcommand{\Xb}{{\bm{X}}}
\newcommand{\be}{{\bm{e}}}

\newcommand{\bI}{{\bm{I}}}

\newcommand{\bbeta}{{\bm{\beta}}}

\newcommand{\bomega}{{\bm \omega}}

\title{BAGEL: A Bayesian Graphical Model for Inferring Drug Effect on Depression Longitudinally in People with HIV}
\author[1]{Yuliang Li}
\author[2]{Yang Ni}
\author[3, 4]{Leah H. Rubin}
\author[5]{Amanda B. Spence}
\author[1, *]{\mbox{Yanxun Xu}}
\affil[1]{Department of Applied Mathematics and Statistics, Johns Hopkins University}
\affil[2]{Department of Statistics, Texas A$\&$M University}
\affil[3]{Departments of Neurology and Psychiatry, Johns Hopkins University School of Medicine}
\affil[4]{Department of Epidemiology, Johns Hopkins Bloomberg School of Public Health}
\affil[5]{\mbox{Department of Medicine, Division of Infectious Disease and Travel Medicine, Georgetown University}}
\affil[*]{Correspondence should be addressed to email: yanxun.xu@jhu.edu}
\date{}                     
\begin{document}

\maketitle

\begin{abstract}
		Access and adherence to antiretroviral therapy (ART) has transformed the face of HIV infection from a fatal to a chronic disease. However, ART is also known for its side effects. Studies have reported that ART is associated with depressive symptomatology. 	
Large-scale HIV clinical databases with individuals'  longitudinal depression records, ART medications, and clinical characteristics 
offer researchers unprecedented opportunities to study the effects of ART drugs on depression over time. 
We develop BAGEL, a Bayesian graphical model to investigate longitudinal effects of ART drugs on a range of depressive symptoms while adjusting for participants' demographic, behavior, and clinical characteristics, and taking into account the heterogeneous population through a Bayesian nonparametric prior. 
We evaluate BAGEL through simulation studies. Application to a dataset from the Women's Interagency HIV Study yields interpretable and clinically useful results. 
BAGEL  not only can improve our understanding of ART drugs effects on  disparate depression symptoms, but also has clinical utility in guiding informed and effective treatment selection to facilitate precision medicine in HIV.

	\noindent \textbf{KEY WORDS:}  Bayesian nonparametrics, Depression, Graphical model, Longitudinal cohort study, Precision medicine. 
	
\end{abstract}

\section{Introduction}
\label{sec:intro}

Antiretroviral therapy (ART) has transformed HIV into a manageable, chronic illness despite its known central nervous system (CNS) side effects, which can complicate disease management. Some of the most commonly reported ART-related CNS adverse effects include depression and anxiety,  suicidal ideation, developmental disorders, and neurological toxicities  \citep{nanni2015depression, zash2018neural}. The presence of these CNS symptoms is a major public health concern as they are associated with ART discontinuation and increases in the likelihood of HIV transmission.  Therefore, a major component of care for people living with HIV includes management and prevention of the long-term adverse effects of ART. 
In this paper, we focus on ART-related  effects on depressive symptomatology.  Depression is one of the leading mental health comorbidities in people with HIV \citep{bengtson2016disparities},  and is associated with poor ART adherence, HIV disease progression, and engagement in risky taking behaviors  \citep{chattopadhyay2017cognitive, ironson2017depression, brickman2017association}.
The high prevalence and harmful effects of depression among people with HIV highlight the need of effective clinical management  and adequate treatment for depression. 

At present, there are more than 30 U.S. Food and Drug Administration-approved ART drugs to treat people with HIV or at risk for HIV. Common ART drugs fall into five drug classes, including nucleotide reverse transcriptase inhibitor (NRTI), non-nucleotide reverse transcriptase inhibitor (NNRTI), protease inhibitor (PI), entry inhibitor (EI), and integrase inhibitor (INSTI). Although ART may alleviate depressive symptoms through suppressing viral load and improving physical health for people with HIV, ART can also exacerbate depressive symptoms through several possible mechanisms, including direct effects on neuronal and mitochondrial functions, astrocyte metabolism, and interference with neurotransmitters  \citep{underwood2015could, shah2016neurotoxicity, cohen2017astrocyte}. 
The ART drugs most commonly associated with depression include efavirenz (EFV; NNRTI) and dolutegravir (DTG; INSTI)  \citep{bengtson2017relationship, elzi2017adverse, borghetti2017efficacy}. 
However, little is known about the effects of most ART drugs and ART combinations on depressive symptoms. Furthermore, ART effects may be confounded by other factors such as sociodemographic, clinical, and behavioral characteristics. Therefore, investigating ART effects on depressive symptoms to optimize health outcomes and facilitate personalized medicine remains a major challenge in HIV studies.

%
%

\begin{figure}[ht!]
	\centering
	\begin{tabular}{cc}
		\includegraphics[width=.5\textwidth]{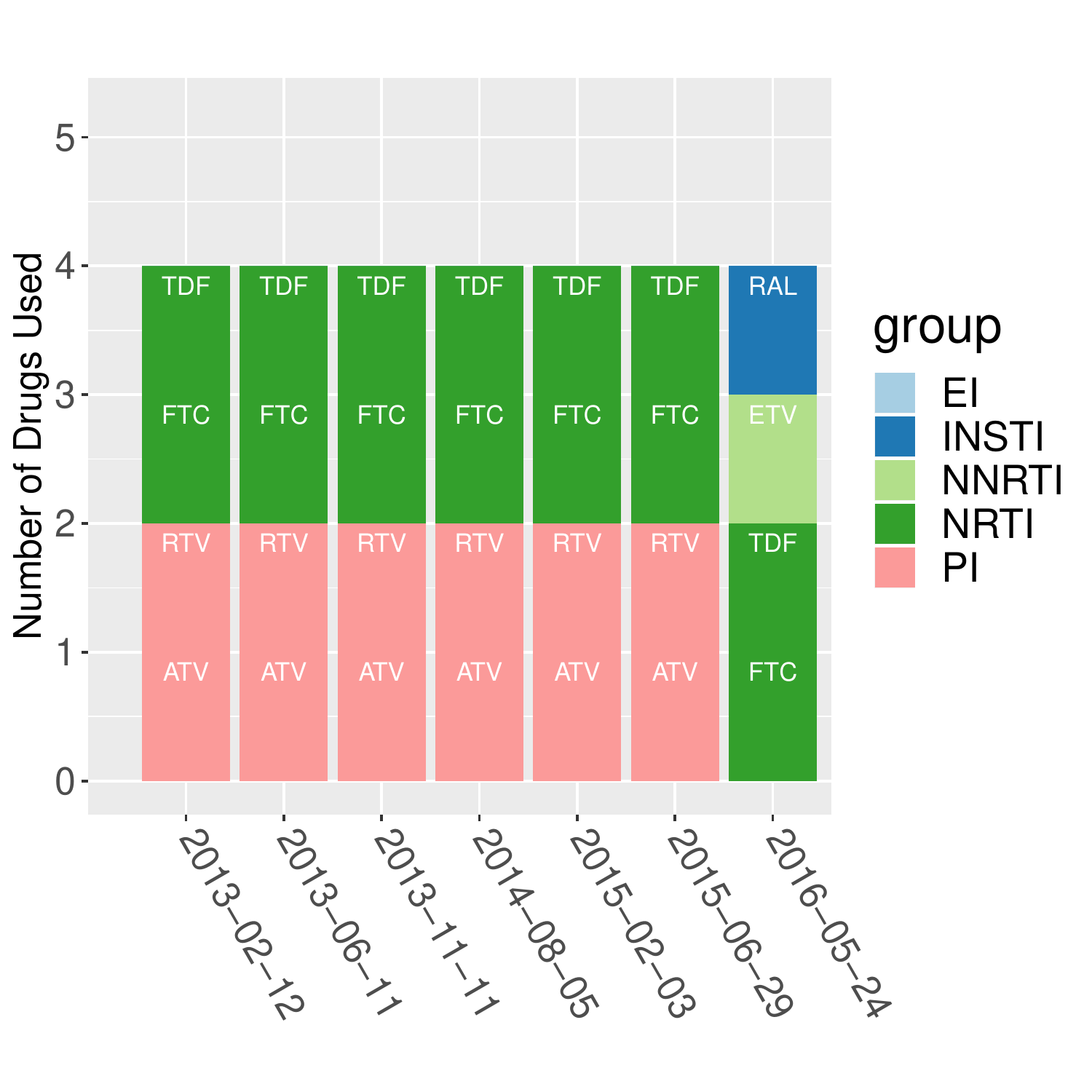}&\includegraphics[width=.5\textwidth]{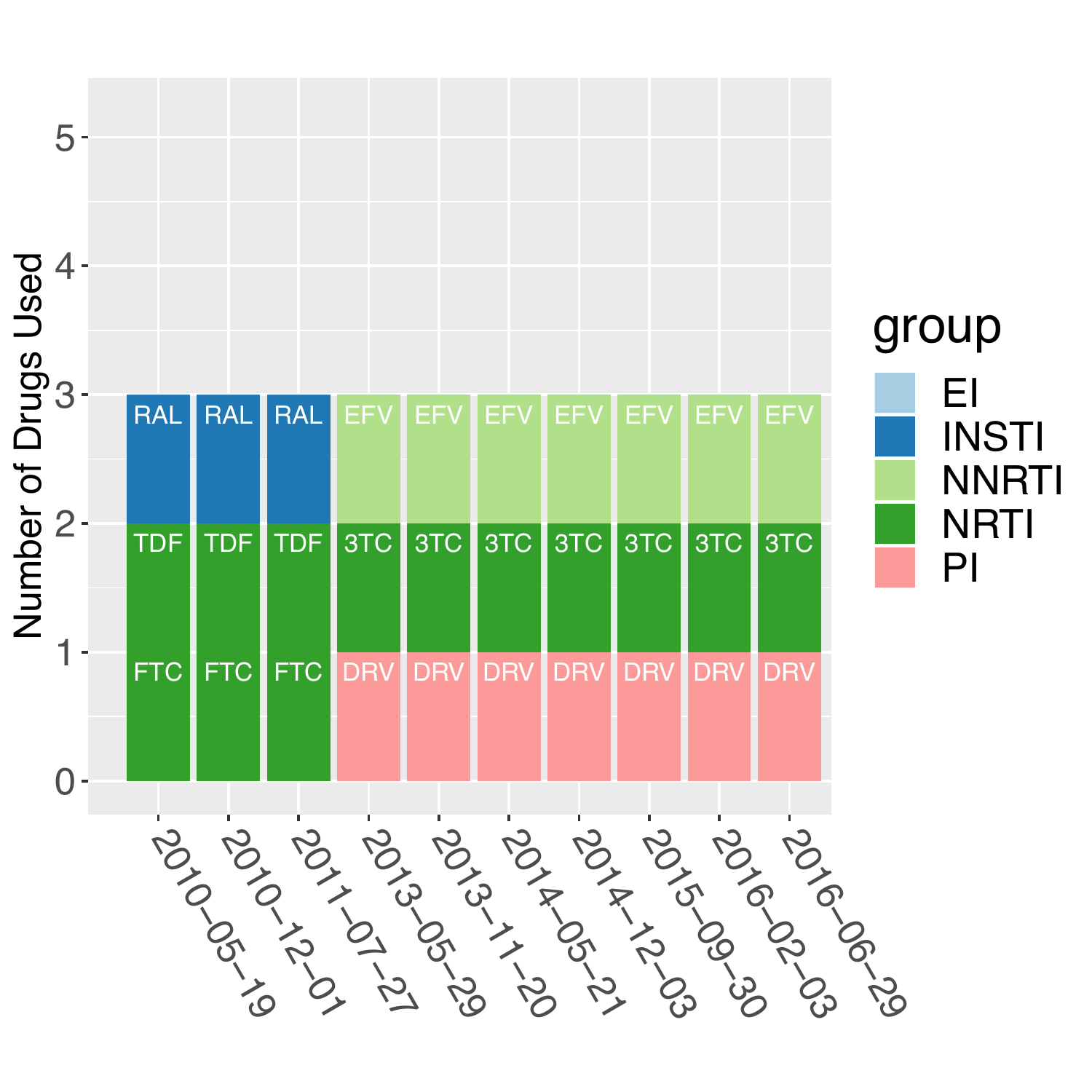}\\
		(a) ART use for participant 1 &(b)ART use for participant  2\\
		\includegraphics[width=.5\textwidth]{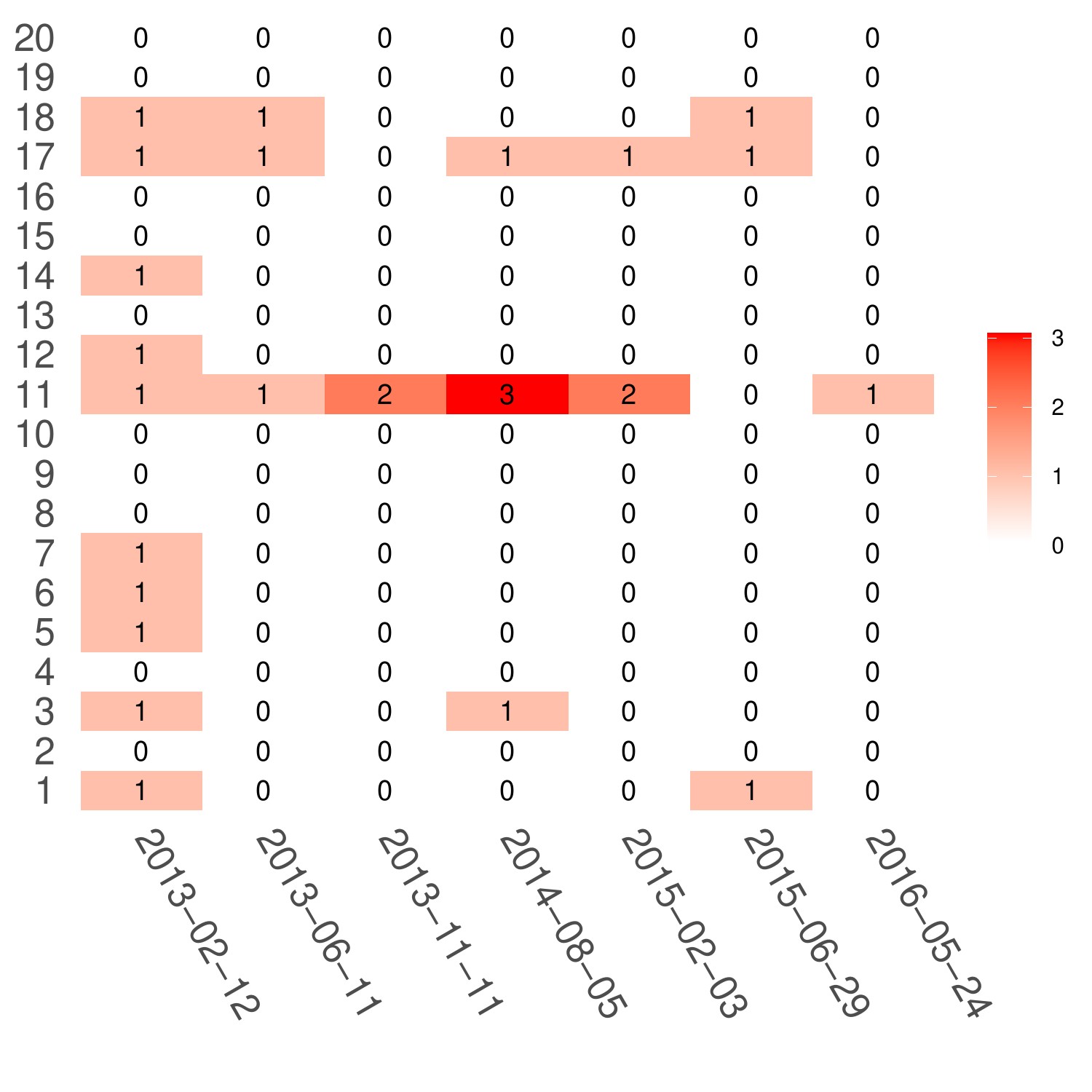}&\includegraphics[width=.5\textwidth]{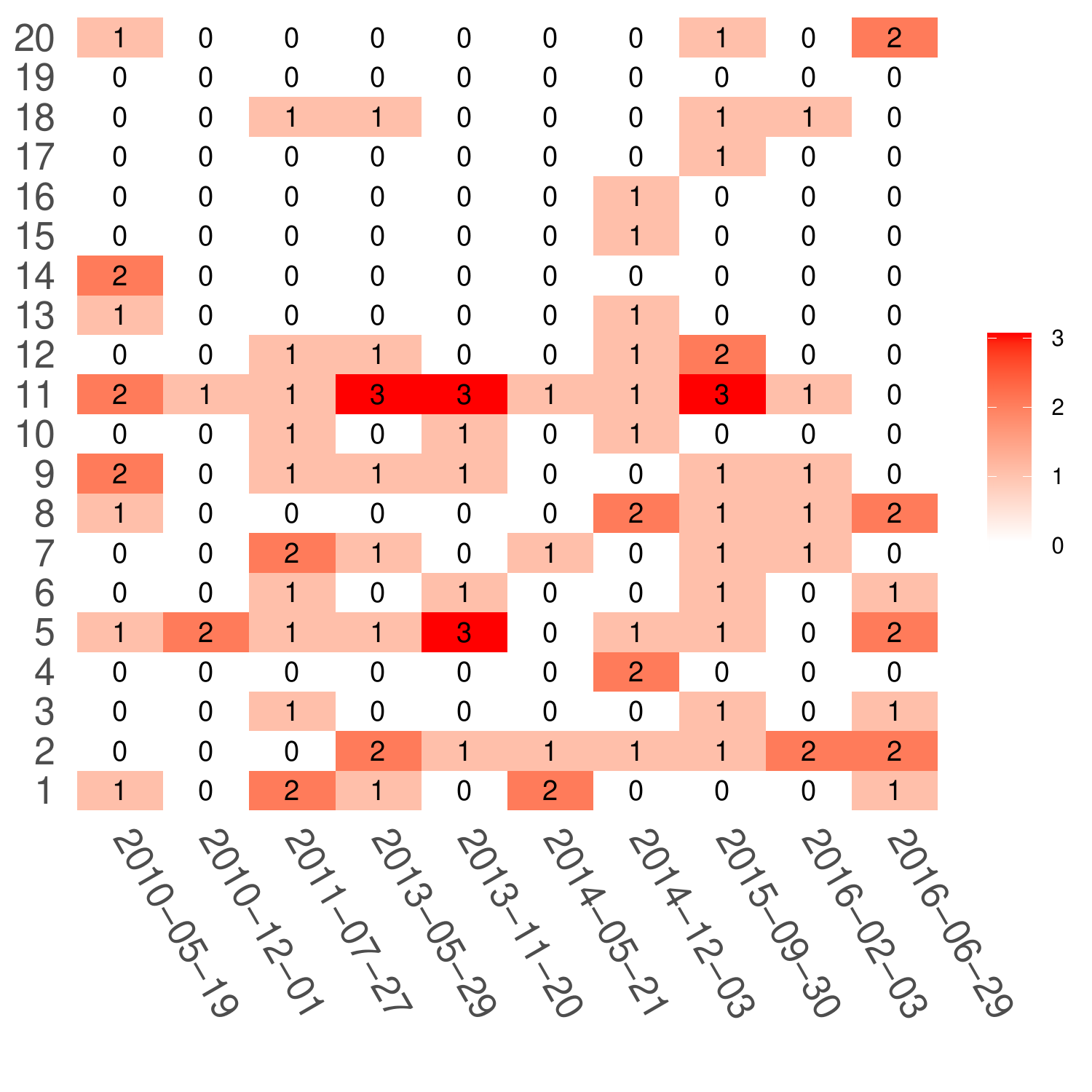}\\
		(a) Depression score for participant 1 &(b) Depression score for participant 2\\
	\end{tabular}
	\caption{(a, b) The number of ART drugs taken by two randomly selected participants from the WIHS versus their visit dates, respectively. Drug names are recorded and different colors represent different drug groups. (c, d) Their depression scores of 20 questions across visits. }
	\label{fig:data}
\end{figure}

The Women's Interagency HIV Study (WIHS) is a large prospective, observational, longitudinal, multicenter study designed to investigate the progression of HIV disease in women with HIV or at risk for HIV infection  in the United States \citep{bacon2005women}.  At semiannual visits,  physical examinations, laboratory testing, and 
questionnaires regarding sociodemographics, medication use, mental health (including depression), and self-reported clinical diagnoses are performed and data is registered in local and national WIHS repositories. Such electronic health records provide us unprecedented opportunities to examine the longitudinal effects of ART drugs on depression after adjusting for socio-demographic, behavioral, and clinical factors. Figure \ref{fig:data}(a, b) present two womens' ART medication history at each of their study visits  (shown as calendar dates). They were followed for different time periods with distinct visit dates and drug usages. At each visit, participants' depressive symptoms were also measured and recorded using the Center for Epidemiologic Studies Depression Scale (CES-D), a 20-item self-administered questionnaire \citep{lewinsohn1997center}; see Figure  \ref{fig:data}(c, d) as an example for the same two participants.  Each question regarding depressive symptoms has an ordinal score in $\{0, 1, 2, 3\}$, with 0 being the least severe and 3 being the most severe. 
The complexity of data including longitudinal laboratory observations, heterogeneous participant population, and dynamic ART assignments, presents analytic and modeling challenges. 


Prior studies examining ART-related effects on depressive symptoms in people with HIV have used a number of self-report instruments to assess depressive symptoms including the 15-item depression subscale of the Hopkins Symptom Checklist (D- HSCL) \citep{derogatis1974hopkins}, the 9-item Patient Health Questionnaire (PHQ-9) \citep{kroenke2001phq}, and the CES-D \citep{lewinsohn1997center}. However, these studies have several limitations. 
First,  instead of considering each individual depressive symptom, these studies use a  sum-score   to represent severity, and a somewhat arbitrary threshold  to distinguish individuals with depression from healthy ones. 
For example, with the CES-D, people with a total score higher than 16 are regarded as having moderate or severe depression. However,  sum-score is an over-simplified measurement for highly heterogeneous diseases like depression. Two
participants with the same score may share no common symptoms or report different symptoms being severe. Moreover, a sum-score assumes equal contributions of all symptoms which are inconsistent with psychometric literature \citep{fried2016measuring}. Therefore, lumping distinct symptoms to a single sum-score and grouping participants with similar sum-scores but different symptoms into one category may result in significant information loss. Second,  as shown in Figure \ref{fig:data},  data in HIV studies is often longitudinal with changing drug assignments and depression scores over the course of treatment,  and the drug effects on depression may change over time and depend on individuals' unique characteristics.
Existing methods only examined the cross-sectional associations between ART drugs and depression. 
Therefore, there is a critical need to develop novel statistical models that are capable of comprehensively examining non-stationary longitudinal effects of all ART drugs on disparate depression symptoms in a heterogeneous HIV population. 

To this end, we develop BAGEL, a Bayesian graphical model  that can simultaneously study   personalized effects of ART drugs on item-level depression symptoms while adjusting for participants' longitudinal covariates, understand the participant-dependent dynamic drug effects, and take into account the heterogeneous HIV population by  clustering participants into subgroups  through a Bayesian nonparametric prior. Specifically, we use a latent bipartite graph with one group of vertices representing ART drugs and the other group of vertices representing depressive symptoms. The presence of an edge between a drug and a depressive symptom indicates a significant drug-depression effect, the size of which is represented by the edge weight. Importantly, the effect sizes can vary across different clinical visits and different participants. Moreover, for broader dissemination and reproducibility, the code implementing BAGEL is available at https://github.com/YanxunXu/BAGEL. 

The rest of paper is organized as follows. In Section \ref{sec:model} we present  modeling details of BAGEL and the posterior inference. The performance of BAGEL is evaluated through simulation studies with comparison to alternative methods in Section \ref{sec:simu}.  In Section \ref{sec:realdata}, we apply BAGEL to a dataset from WIHS to study the  longitudinal ART drugs effects on depression items and demonstrate the clinical utility of BAGEL. 
We conclude with a brief discussion in Section \ref{sec:discuss}.

\section{Model and Inference}
\label{sec:model}

\subsection{Probability model}
Let $U_{ijq} \in \{0, 1, 2, 3\}$ be the ordinal score of depression item $q$ for participant $i$ at visit $j$, $i=1, \dots, n$, $j=1, \dots, J_i$, and $q=1, \dots, Q$. In WIHS, we have $Q=20$ depression items, which measure different depressive symtoms such as ``bothered by things", ``appetite was poor", and ``fearful." A detailed description of all  20 symptoms  is provided in Section  \ref{sec:realdata}. A higher score indicates a worse depressive symptom. 
Following \cite{albert1993bayesian}, we introduce a latent continuous random variable $Y_{ijq}$ such that $U_{ijq}=k$ if and only if $Y_{ijq}\in (a_k, a_{k+1}]$, where $k=0, \dots, 3$, $a_0=-\infty$, and $a_4=\infty$. For identifiability, we set $a_1=0$. Let $Z_{ijd}$ be a binary indicator to represent whether participant $i$ at visit $j$ uses drug $d$, $d=1, \dots, D$, and denote $\bZ_{ij}=(Z_{ij1}, \dots, Z_{ijD})$. Let $\bX_{ij}$ be an $S$-dimensional row vector including an intercept, time-invariant covariates (e.g., race), and time-varying covariates (e.g., BMI, CD4 count).

We model the latent continuous depression scores $\bY_{ij}=(Y_{ij1}, \dots, Y_{ijQ})$ as follows, 
\begin{eqnarray}
\bY_{ij} =\bX_{ij}\bbeta_{i}+  \bZ_{ij}\bB_{ij}+ \bomega_{ij} +\beps_{ij}, \ \ i=1,\dots n,~~ j=1, \dots, J_i,
\label{eq:mainmodel}
\end{eqnarray}
with $\beps_{ij}=(\eps_{ij1}, \dots, \eps_{ijQ})$. 
The first term $\bX_{ij}\bbeta_i$ captures the dependence of the latent outcome $\bY_{ij}$ on the covariates $\bX_{ij}$, where $\bbeta_{i}$ is an $S\times Q$ matrix representing the covariate effects for participant $i$. 
The second term $\bZ_{ij}\bB_{ij}$ models the drug effects where $\bB_{ij}$ is a $D\times Q$ matrix with each element $B_{ij, dq}$ being the contribution of drug $d$ to depression item $q$ for participant $i$ at visit $j$; the modeling details of drug effects $\bB_{ij}$ will be introduced later.  
By study design, depression items are correlated. For example, in our data, the pairwise rank correlations of observed depression ordinal scores range from 0.11 to 0.66. 
To capture such dependencies, we assume $\bomega_{ij}\in \mathbb{R}^Q$ follows a centered multivariate normal distribution $\mathrm{MN}({\bm 0},  \sigma^2_{\epsilon}\bm{C}_{\omega})$.	The sampling model is completed by assuming independent normal errors $\beps_{ij}\sim \mathrm{MN}({\bm 0}, \sigma^2_{\epsilon}{\bm I})$, where $\sigma_\epsilon^2=1$ for identifiability.

\begin{figure}[ht!]
	\centering
	\includegraphics[width=.85\textwidth]{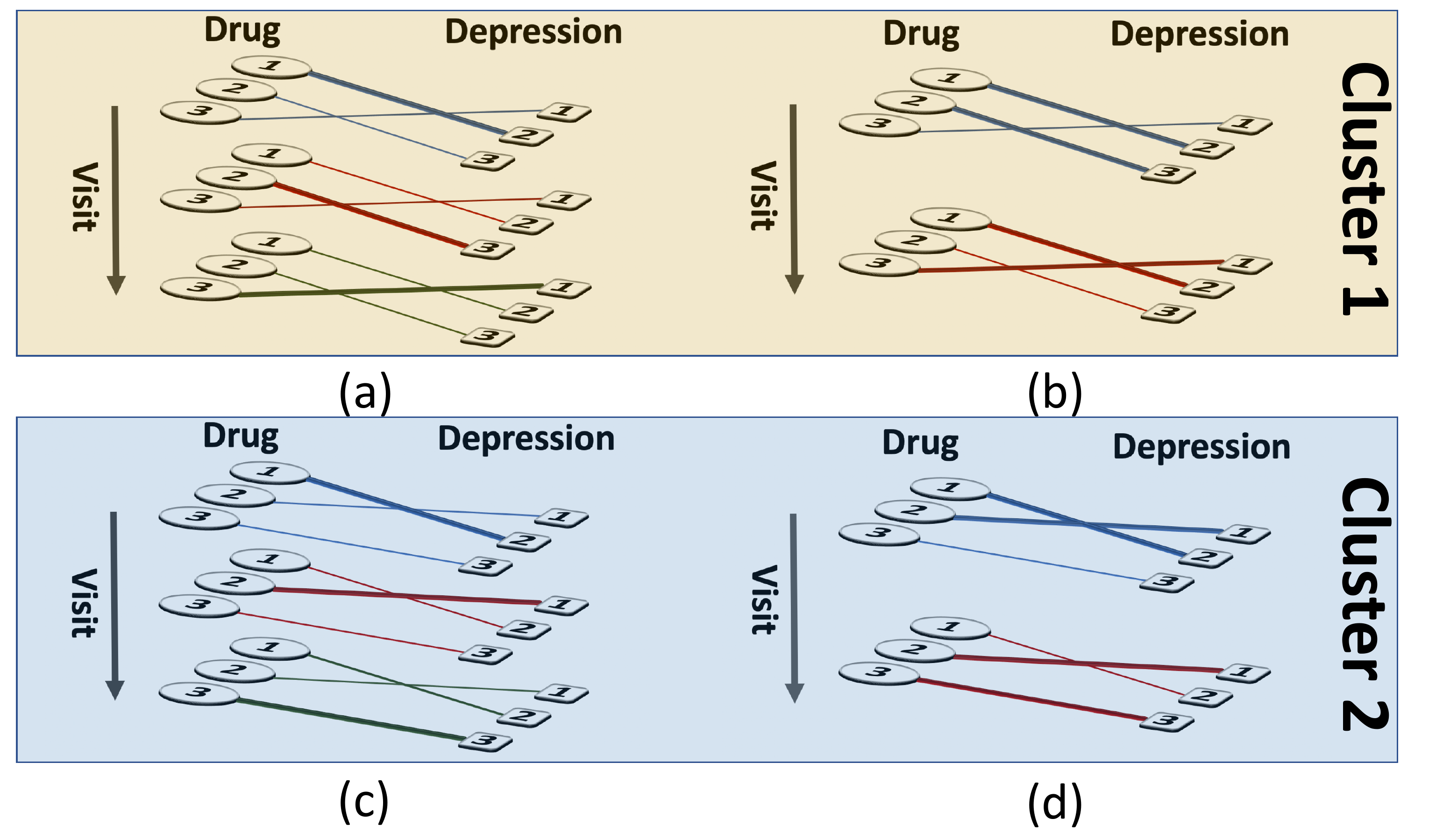}
	\caption{Drug-depression relationships across visits for four participants (a)-(d) in two different clusters. Different colors encode different visits for one participant, and the width of each edge represents the weight of the association between the corresponding drug (circles) and depression (squares). }
	\label{fig:bipartite}
\end{figure}

\noindent{\textbf{Modeling drug-depression effects.}} The main contribution of the proposed model is the  formulation of the drug effects $\bB_{ij}$, which are the key parameters of interest. Since there are 20 depression items and more than 20 drugs, to encourage parsimony, we assume $\bB_{ij}$ to be a sparse matrix, i.e., not all drugs are associated with every depression item. Letting $\circ$ denote the Hadamard   product of two matrices of the same dimension, we decompose $\bB_{ij}=\bR_i\circ \bLambda_{ij}$ multiplicatively into two components. The first component $\bR_i=(R_{i,dq})$ is a $D\times Q$  binary matrix that induces sparsity in $\bB_{ij}$ with $R_{i,dq}=1$ if drug $d$ is significantly associated with depression item $q$ for participant $i$ and $R_{i,dq}=0$ otherwise. The matrix $\bR_i$ can be also viewed as a directed acyclic graph adjacency matrix of a bipartite network consisting of two sets of nodes: drugs and depression items; see Figure \ref{fig:bipartite} for example.
The second component $\bLambda_{ij}$  is a $D\times Q$ matrix that quantifies the strength of the (non-zero) drug-depression associations over time with each element $\Lambda_{ij, dq}$ being the effect of drug $d$ on depression item $q$ for participant $i$ at visit $j$. 
Note that although the presence or absence of the drug-depression links $\bR_i$ for participant $i$ do not change over time, we allow the strength of the links $\bLambda_{ij}$ to vary with time and covariates through the following model,
\begin{eqnarray}
\Lambda_{ij, dq} = \widetilde{\bX}_{ij} \balpha_{i, dq} + s(t_{ij}), 
\label{eq:lambda}
\end{eqnarray}
where $ \widetilde{\bX}_{ij} \balpha_{i, dq}$ accounts for the dependence of the drug effect on participants' covariates $\widetilde{\bX}_{ij}$, 
$t_{ij}$ denotes the time of visit $j$ since the first visit for participant $i$, and $s(\cdot)$ is a nonlinear smooth function. In general, $\widetilde{\bX}_{ij}$ could be different from $\bX_{ij}$ but in our later simulations and application, $\widetilde{\bX}_{ij}=\bX_{ij}$.

We use splines to model the nonlinear function $s(\cdot)$ by letting $s(t_{ij})=\tilde{\bt}_{ij}\bm{\gamma}_{i,dq}$, where $\tilde{\bt}_{ij}=(\tilde{t}_{ij1},\dots,\tilde{t}_{ijB})$ denotes the cubic B-spline basis expansion of $t_{ij}$ and $B$ is the number of bases. 
To prevent overfitting, we leverage the P-spline \citep{eilers1996flexible} that penalizes 
finite differences of adjacent B-spline coefficients to encourage smoothness.
%
As shown in \cite{lang2004bayesian}, P-splines can be written as a Bayesian hierarchical model by assuming $\bm{\gamma}_{i,dq}\sim \mathrm{MN}(0, \sigma^2_{\gamma}{\bm K}^{-})$, where ${\bm K}$ is a singular penalty matrix constructed from the second-order differences of the adjacent spline coefficients and ${\bm K}^{-}$   is the pseudo-inverse of the matrix ${\bm K}$.  
In summary, the term $\bZ_{ij}\bB_{ij}$ represents (i) the drug main effect, (ii) drug-covariate interaction, and (iii) nonlinear drug-time interaction. 

\subsection{Priors}
The proposed model is parameterized by $(\{\Theta_i\}_{i=1}^n, \bm{C}_{\omega})$, where 
$\Theta_i=(\bbeta_i, \bR_i, \{\balpha_{i, dq}, \bm{\gamma}_{i, dq}\}_{d=1, q=1}^{D, Q})$. 
Although one can apply model \eqref{eq:mainmodel} independently to one participant at a time,  it is deemed inefficient and hard to interpret, especially for individuals with few visits. We propose a joint modeling approach through a Bayesian nonparametric (BNP) prior on $\Theta_i$'s to  borrow information across participants for estimating personalized drug-depression relationships over time.  
We consider a Dirichlet Process (DP) prior \citep{ferguson1974prior}, $\Theta_i\iidsim G$ and $G\sim DP(m_0,G_0)$ with a concentration parameter $m_0$ and a baseline measure $G_0(\cdot)$. DP has been widely used in various biomedical applications \citep{muller2004nonparametric, xu2016non} to account for sampling heterogeneity.  DP prior gives rise to ties of $\Theta_i$'s due to its  almost surely discreteness, which elegantly defines a natural partition of participants without a pre-specified number of clusters --- $i$ and $i'$ belong to the same cluster if $\Theta_i=\Theta_{i'}$. Because of the presence of ties, let $\{\tTheta_h\}_{h=1}^H$ with $\tTheta_h=(\tilde{\bbeta}_h, \tilde{\bR}_h, \{\tilde{\balpha}_{h, dq}, \tilde{\bm{\gamma}}_{h, dq}\}_{d=1, q=1}^{D, Q})$ denote the unique values of $\{\Theta_i\}_{i=1}^n$. Let $e_i$ denote the clustering membership indicator such that $e_i=h$ indicates participant $i$ belongs to cluster $h$. Then $\Theta_i=\tTheta_h$ if $e_i=h$. Given $e_i$, the DP prior induces the following prior on $\Theta_i$, 
$$\Theta_i = \sum_{h=1}^H \tTheta_h I(e_i=h) \mbox{~~and~~} \tTheta_h\sim G_0.$$
In essence, the DP prior allows for clustering of participants by grouping participant-specific parameters $\{\Theta_i\}$ into $\{\tTheta_h\}$ and makes inference for cluster-specific parameters $\tTheta_h$ using participants within cluster $h$. Figure \ref{fig:bipartite}(a-d) illustrates the drug-depression relationships for four participants in two different clusters with three visits recorded for participants (a) and (c) and two visits recorded for participants (b) and (d). Participants belonging to the same cluster share the same graph structure (encoded in $\bR_i$), although the edge weights can differ across participants and visits (quantified by $\bB_{ij}$). 
Furthermore, the BNP prior on $\Theta_i$ yields a flexible nonparametric response surface for $\bY_{ij}$ capturing complex drug effect dynamics. 

Let $\tilde{\bbeta}_{hq}$  be the $q$-th column of $\tilde{\bbeta}_h$. We assume a conjugate base measure $G_0$ of DP, $G_0(\tTheta_h)=\prod_{q=1}^Q p(\tilde{\bbeta}_{hq}) \prod_{d=1,q=1}^{D, Q}p(\tilde{R}_{h, dq})p(\tilde{\bm{\alpha}}_{h, dq})p(\tilde{\bm{\gamma}}_{h, dq})$ where $p(\tilde{\bbeta}_{hq})=\mathrm{MN}(0, \sigma^2_{\beta}\bI)$, $p(\tilde{\bm{\alpha}}_{h, dq})=\mathrm{MN}(0, \sigma^2_{\alpha}\bI)$, $p(\tilde{\bm{\gamma}}_{h, dq})=\mathrm{MN}(0,  \sigma^2_{\gamma}{\bm K}^{-})$, and $p(\tilde{R}_{h, dq})=\text{Bernoulli}(\rho)$. We consider a non-informative $\text{Unif}(-1,1)$ prior for the off-diagonal elements of $\bm{C}_{\omega}$, although
alternative priors could be adopted if certain prior knowledge on the correlation structure of depression items is available.  We complete the model construction by assigning the following hyperpriors:  $\sigma_\beta^2\sim \text{inv-Gamma}(a_\beta, b_\beta)$, $\sigma_\alpha^2\sim \text{inv-Gamma}(a_\alpha, b_\alpha)$, $\sigma_\gamma^2\sim \text{inv-Gamma}(a_\gamma, b_\gamma)$, and $\rho\sim\text{Beta}(\alpha_{\rho}, \beta_{\rho})$.  The posterior inference is carried out using Markov chain Monte Carlo (MCMC) simulations. Details of the sampling procedure are described in Section A of the Supplementary Material.  

%
%


%

\section{Simulation Study}
\label{sec:simu}
In this section, we evaluated the performance of BAGEL through simulation studies by comparing posterior inference with simulation truth. Furthermore, to demonstrate the advantages of (i) imposing a DP prior to account for participants' heterogeneity and (ii) encouraging sparsity in the drug-depression effects $\bB_{ij}$, we compared BAGEL to two alternative methods. The first method  does not take into account participants' heterogeneity by assuming that all participants shared the same parameter, i.e.,  $\Theta_i=\Theta$ and $\Theta\sim G_0$. We call this method Homogeneous. The second method, called NoSparsity,  dropped the sparse binary matrix $\bR_i$ from BAGEL, i.e., assuming $\bB_{ij}=\bLambda_{ij}$.

\subsection{Simulation setup}
We considered two scenarios, one with $n=200$ participants and the other with $n=500$ participants.  In both scenarios, we assumed that there were $D=5$ drugs and $Q=3$ depression items, with the true number of clusters being $H_0=3$. All participants were randomized to the three clusters with equal probabilities. For each participant $i$, we generated the number of visits $J_i$ from a Poisson distribution with mean 10. The duration between two consecutive visits (i.e., $t_{i,j+1}-t_{ij}$) was generated from a normal distribution with the mean 1 year and standard deviation 0.2. Let $\mbox{N}(\mu,\sigma^2)$ denote a normal distribution with mean $\mu$ and variance $\sigma^2$. For participants' covariates $\bX_{ij}$, we considered two time-invariant covariates with one being generated from $\mathrm{Bernoulli}(0.5)$ and the other being generated from $\mbox{N}(0, 4)$, and three time-variant covariates mimicking age, body mass index (BMI), and smoking status. We randomly sampled participants' ages at their initial visits from the motivating WIHS dataset with replacement, then computed their ages at followup visits based on the generated $t_{ij}$'s. The BMI and smoking status were generated  from $\mbox{N}(28, 8)$ and $\mathrm{Bernoulli}(0.6)$  independently across participants and their visits, respectively.  
The drug usage $Z_{ijd}$ for participant $i$ at visit $j$ was generated from $\mathrm{Bernoulli}(0.8)$ independently.

Conditional on clustering memberships, the simulated true values of $\bbeta_i$'s and $\balpha_{i}$'s in both scenarios are tabulated in Supplementary Tables T1 and T2. For the matrix $\bR_i$ that induces sparsity in drug-depression effects, we generated each element $R_{i, dq}$ independently from $\mathrm{Bernoulli}(0.4)$. The simulated true $s(t_{ij})$ was assumed to be $s(t_{ij}) = 6 t_{ij}$.  Setting $\bm{C}_{\omega}$ to be $\begin{bmatrix}
1 & 0.3 & 0.35 \\
0.3 & 1 & 0.4 \\
0.35 & 0.4 & 1
\end{bmatrix}$ and the threshold to be $(a_1, a_2, a_3)=(0, 8, 16)$, we generated the latent continuous $Y_{ijq}$ from \eqref{eq:mainmodel} as well as the corresponding ordinal depression scores $U_{ijq}$, $i=1, \dots, n; j=1, \dots, J_i; q=1, \dots, Q. $

We applied BAGEL to the simulated dataset with 100 repeated simulations in both scenarios. The hyperparameters were set to be $m_0 = 1, a_{\beta} = a_{\alpha} = a_{\gamma} = 3, b_{\beta} = b_{\alpha} = b_{\gamma} = 10, \alpha_\rho = 1, \text{ and } \beta_\rho = 10$. In each analysis, we ran 20,000 MCMC iterations with an initial burn-in of 10,000 iterations, thinned by 10. We assessed the convergence using R package {\it coda}, showing no sign of lack of convergence.

\subsection{Simulation results}
We first evaluated the clustering performance of BAGEL. BAGEL successfully identified $\hat{H}=3$ clusters in both scenarios, since the marginal posterior probabilities of the simulated true number of clusters averaged over 100 repeated simulations were $p(H=3\mid \mathrm{data})=99.6\%$ when $n=200$, and $p(H=3\mid \mathrm{data})=89.1\%$ when $n=500$. We calculated the posterior  co-clustering probabilities of participants based on the empirical proportion of participants being clustered together over the post-burn-in MCMC samples. The simulated true clustering schemes and the co-clustering probability matrices for one randomly selected simulation under both scenarios are depicted in Supplementary Figure F1, showing that BAGEL assigns participants to their simulated true clusters with high probabilities.

Reporting cluster-specific parameters is challenging due to the issue of label switching. Following \cite{dahl2006model},   
let $e_i$ denote the clustering membership indicator of participant $i$ and $\mathbf{V}$ be an $n\times n$ matrix whose entry $V_{i_1, i_2} = \mathbb{P}(e_{i_1} = e_{i_2})$ represents the posterior probability of participants $i_1$ and $i_2$ being clustered together, which can be computed based on posterior samples.  In each MCMC iteration, we computed an $n\times n$ binary matrix $\bold{V}^{\be}$, whose entry $V^{\be}_{i_1,i_2} = \mathbb{I} (e_{i_1} = e_{i_2})$ indicates whether participants $i_1$ and $i_2$ are clustered together at that iteration.  
The clustering  that minimizes the Frobenius distance between $\bold{V}^{\be}$ and $\bold{V}$, given by $\hat{\be}=\text{arg min}_{\be}\|\bold{V}^{\be}- \bold{V}\|$, is called the  least-square summary of clustering. Then we relabeled the cluster memberships at each MCMC iteration by minimizing its Frobenius distance to the least-square summary as a post-processing step \citep{li2019bareb}.  Supplementary Figures F2 and F3 plot  posterior means and 95\% credible intervals (CIs) for $\tilde{\bm \beta}_{hq}$'s and $\tilde{\bm \alpha}_{h, dq}$'s averaged over 100 repeated simulations, respectively, showing that all 95\% CIs are centered around the simulated true values. To assess the ability of BAGEL in recovering the sparsity ${\bm R}_i$ of drug effects, we plotted the posterior probabilities of $\tilde{R}_{h, dq}$ being equal to the simulated true values in Supplementary Figure F4 under both scenarios, indicating satisfactory performance. We also computed the true positive rates (TPR) and false discovery rates (FDR) for recovering ${\bm R}_i$ under both scenarios: when $n=200$, TPR=0.96, FDR=0.08; when $n=500$, TPR=0.98, FDR=0.02,  indicating good recovery.

\begin{figure}[ht!]
	\begin{centering}
		\begin{tabular}{ccc}
			\includegraphics[width=0.3\textwidth]{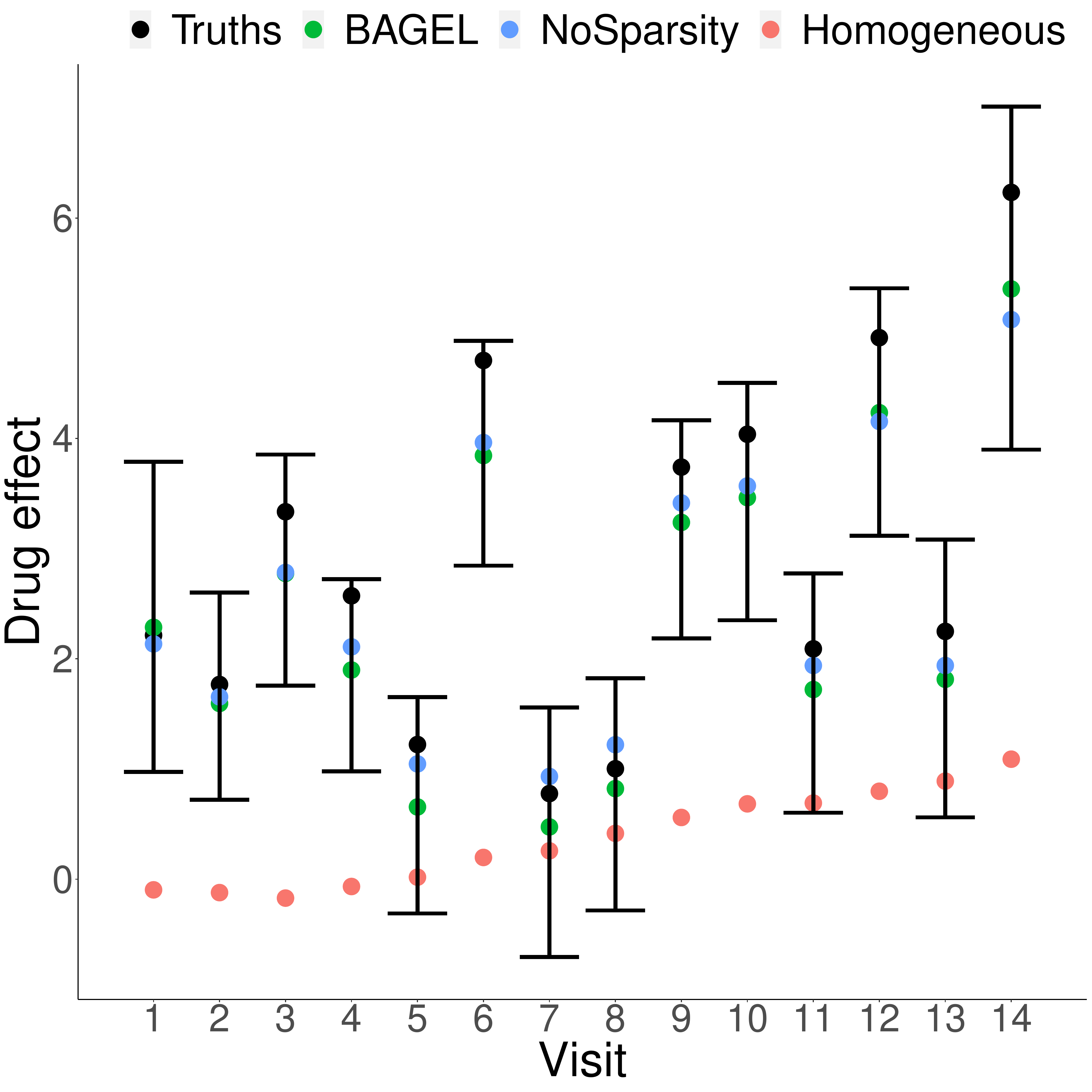}&\includegraphics[width=0.3\textwidth]{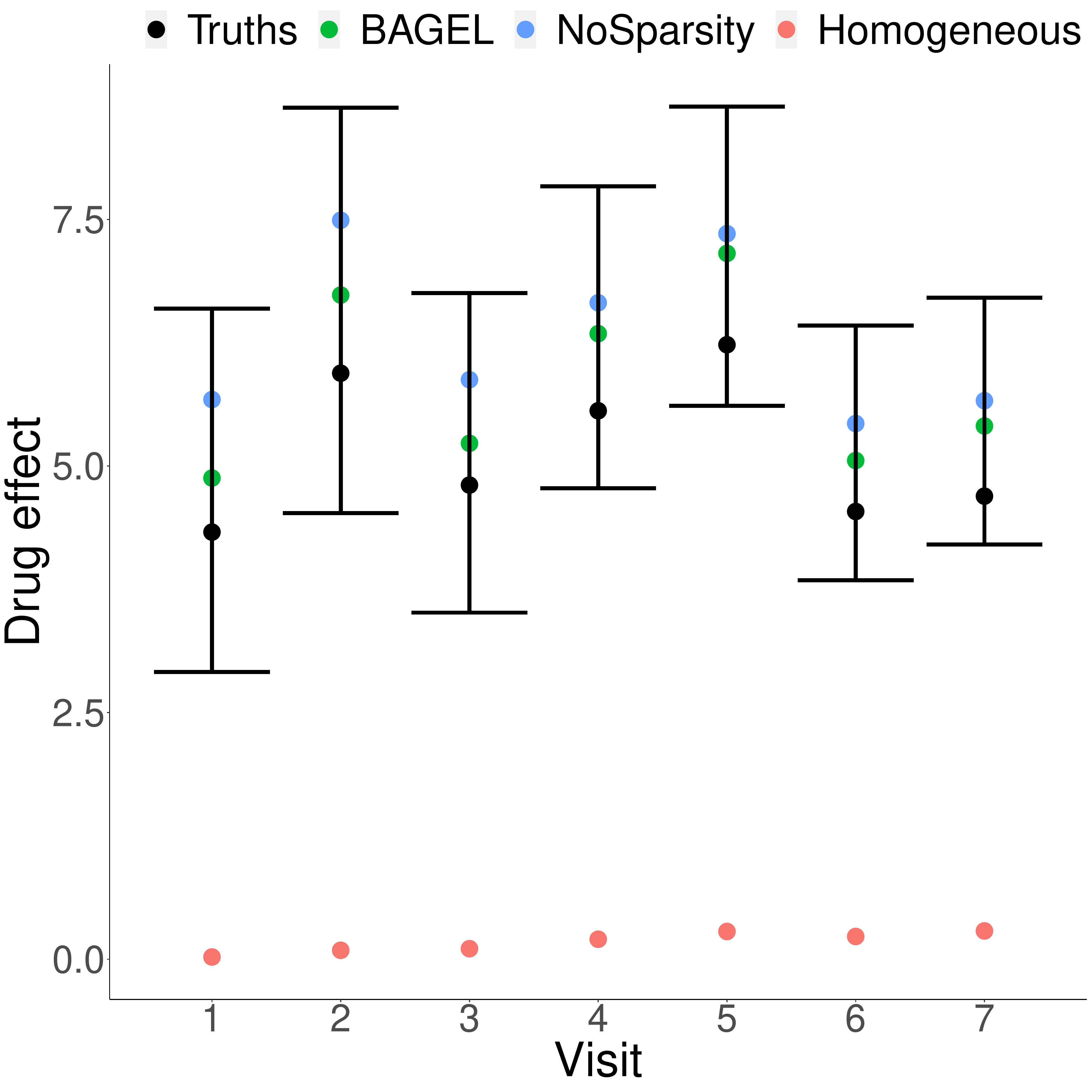}&\includegraphics[width=0.3\textwidth]{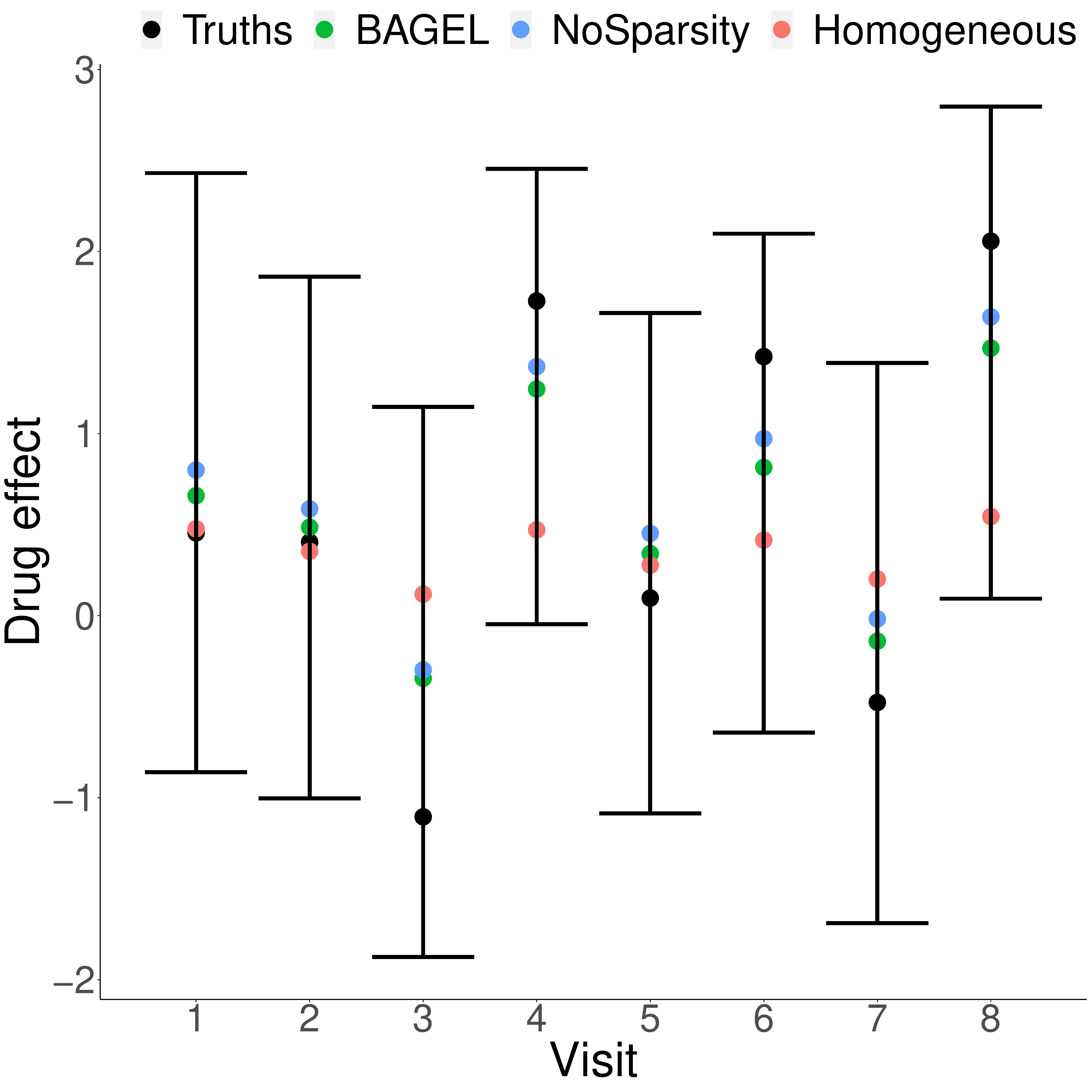}\\
			(a) participant in cluster 1 &(b) participant in cluster 2 &(c) participant in cluster 3
		\end{tabular}
	\end{centering}
	\caption{Simulated true values, posterior means with 95\% CIs of the estimated $B_{ij, dq}$ under BAGEL, and posterior means under two alternative methods: NoSparsity and Homogeneous for three participants randomly selected from each of the three clusters, respectively.  
		The black dots represent simulation truths; the green dots represent posterior estimates under BAGEL; the blue and pink dots represent posterior estimates under NoSparsity and Homogeneous, respectively.}
	\label{fig:bij}
\end{figure}

Next, we examine whether BAGEL can recover individualized longitudinal drug effects $\bB_{ij}$, $j=1, \dots, J_i$. We randomly selected one participant from  each of the three clusters in one simulated dataset when $n=200$. These participants had 14, 7, and 8 visits, respectively. 
As drug effects  for one participant are different across different drugs and depression items, we computed the posterior summaries of $B_{ij, dq}$ for the randomly selected $d$ and $q$ such that $R_{i,dq}\neq 0$ for illustration (note that when $R_{i, dq}=0$, the corresponding $B_{ij, dq}=0$). 
Figure \ref{fig:bij} plots the estimated posterior means of $B_{ij, dq}$'s with 95\% CIs for these three participants, indicating that BAGEL can successfully recover the simulated true drug effects. 

Furthermore, we compared BAGEL with two alternative methods: Homogeneous and NoSparsity. Figure \ref{fig:bij} shows the estimated posterior means of $B_{ij, dq}$ under the two alternatives. Both BAGEL and NoSparsity estimated the drug effects much better than the Homogeneous method, especially for participants from clusters 1 and 2 partly because the drug effects may be averaged out in a heterogeneous population and hence greatly biased towards zero. BAGEL slightly outperformed NoSparsity with substantially better interpretability due to sparsity. In addition, as shown in Supplementary Figures F5 and F6, the CIs for the model parameters of NoSparsity tend to be wider than those of BAGEL. We also compared the three methods in terms of Watanabe-Akaike information criterion (WAIC) \citep{watanabe2010asymptotic,gelman2014understanding}. BAGEL yielded the smallest WAIC, which was 3688.73, while the WAIC values of NoSparsity and Homogeneous were 3816.89 and 9602.72, respectively.



%

\section{The WIHS Data Analysis}
\label{sec:realdata}
\subsection{Data}
The WIHS is a multi-center, longitudinal study developed to characterize the   natural and treated history of HIV infection and clinical outcomes in women residing in the United States
\citep{barkan1998women, bacon2005women,adimora2018cohort}.  WIHS participants complete ``core" visits approximately every 6 months. At each visit, women undergo 
clinical examination,  structured medical and psychosocial interviews, and laboratory testing to assess HIV status/viral load.  For the present analysis, we included all visits after January 2012 from women in the Washington, D.C. WIHS site resulting in a total of 214 participants with 1,240 visits. 
The primary outcome of interest was the item-level responses on CES-D, a 20-item self-administered questionnaire, which is 
commonly used to assess depression in HIV studies \citep{moore1999severe,ickovics2001mortality, maki2012depressive}.  The CES-D measures the frequency of depressive symptoms (e.g., ``people dislike me") during the  week prior to the visit, where ``0" indicates no symptom or the duration of symptom is less than one day, ``1" indicates the duration of symptom is between one  and two days, ``2" indicates the duration of symptom is between three and four days, and ``3" indicates the duration of symptom is longer than five days.  
Figure \ref{fig:edge} lists all 20 symptoms, which can be categorized in three types: somatic, affect, and interpersonal symptoms.   Somatic symptoms include unpleasant or worrisome mood such as ``restless", ``appetite", and ``concentration." 
Affect symptoms include the lack of positive affect (e.g., ``enjoyed life", ``hopeful of future") and/or the presence of negative affect (``fearful", ``lonely", ``failure").  Interpersonal symptoms include``people disliked me" and ``people  unfriendly". 
The items reflecting positive affect were reversed scored so that higher values on each item reflected more negative symptoms.

We included the following sociodemographic, behavioral, and clinical covariates as risk factors \citep{cook2002effects, rubin2011perinatal, maki2012depressive} for depressive symptoms: age, body mass index (BMI), CD4 count, nadir CD4 count (CD4 Nadir), viral load (VLOAD), race, smoking, substance abuse (e.g., marijuana, cocaine, and heroin), and education. Table \ref{tab:covariates} summarizes these characteristics of participants at their first visits in the dataset. There were 23 drugs used in the dataset, falling into five drug classes: NRTI, NNRTI, PI, EI, and INSTI. The drug use frequency varies significantly among different drugs from hundreds of visits to only few visits, the details of which are reported in Supplementary Table T3. 


\begin{table}
	\begin{centering}
		{\small
			\hspace{1cm}
			\begin{tabular}{p{5cm}ccccc} 
				\makebox[3cm][l]{   }        & Overall &  \multicolumn{4}{c}{Cluster}\\ \cline{3-6}
				\makebox[3cm][l]{   } & (n=199) &  1 (n=94) &  2 (n=105) & 3 (n=6) & 4 (n=9)  \\
				\makebox[3cm][l]{  Variables } &  n(\%) &  n(\%) &  n(\%) &  n(\%) &  n(\%)  \\
				\hline
				\rowcolor{Lavender} \multicolumn{6}{l}{Demographics} \\
				\makebox[3cm][l]{Age (years)} & & &  & & \\
				\makebox[3cm][r]{$\leq35$} & 10 (5) & 4 (4)     & 6 (6) & 0 (0)  & 0 (0)\\
				\makebox[3cm][r]{36-45}     & 66 (33) & 28 (30) & 38 (36) & 3 (50) & 5 (56)\\
				\makebox[3cm][r]{46-55}     & 80 (40) & 40 (43) & 40 (38) & 2 (33) & 3 (33)\\
				\makebox[3cm][r]{$>55$}    & 43 (22) & 22 (23)& 21 (20) & 1 (17) & 1 (11)\\
				
				\makebox[3cm][l]{BMI} & & &   & & \\
				\makebox[3cm][r]{$<18.5$}   & 6 (3)    &3 (3)     &  3 (3)        & 0 (0)  & 0 (0)\\
				\makebox[3cm][r]{18.5-29.9} & 107 (54) &50 (53) & 57 (54)  & 5 (83) & 4 (44)\\
				\makebox[3cm][r]{30-39.9}    & 59 (30) & 25 (27)   & 34 (32) & 1 (17)  & 4 (44) \\
				\makebox[3cm][r]{$\geq 40$} & 27 (13) & 16 (17)    & 11 (11) & 0 (0)  & 1 (12)  \\
				
				\makebox[3cm][l]{Race}   & & &  & &  \\
				\makebox[3cm][r]{White}  & 35 (18) & 12 (13)    & 24 (23)  & 0 (0)  & 0 (0)   \\
				\makebox[3cm][r]{Black}  & 150 (75) & 76 (81)    & 74 (70) & 6 (100) & 8 (89)\\
				\makebox[3cm][r]{Others} & 13 (7) & 6 (6)      & 7 (7)          & 0 (0) & 1 (11)\\
				
				\rowcolor{Lavender} \multicolumn{6}{l}{HIV-related clinical characteristics} \\
				\makebox[3cm][l]{CD4} & & &  &  &  \\
				\makebox[3cm][r]{$\leq250$}     & 18 (9)    & 10 (11)     & 8 (8)     & 1 (17) & 0 (0)  \\
				\makebox[3cm][r]{251-500}        & 57 (29)   & 32 (34)    & 25 (24)  & 2 (33) & 3 (33) \\
				\makebox[3cm][r]{501-1000}      & 107 (54) & 44 (47)     & 63 (60)  & 3 (50) & 6 (67) \\
				\makebox[3cm][r]{$\geq 1001$} & 17 (8)     & 8 (8)          & 9 (8)  & 0 (0) & 0 (0) \\
				
				\makebox[3cm][l]{CD4NADIR} & & &  &  &  \\
				\makebox[3cm][r]{$\leq250$}     & 74 (37)  & 31 (33)& 43 (41) & 2 (33) & 3 (33)  \\
				\makebox[3cm][r]{251-500}        & 98 (49) & 48 (51) & 50 (48) & 4 (67) & 4 (45) \\
				\makebox[3cm][r]{$> 500$}            & 27 (14)   & 15 (16) &  12 (11) & 0 (0)  & 2 (22)\\
				
				\makebox[3cm][l]{VLOAD} & & &  & &  \\
				\makebox[3cm][r]{$\leq500$}       & 168 (85) & 81 (87) & 87 (83)  & 5 (83) & 8 (89) \\
				\makebox[3cm][r]{501-5000}        & 14 (7)    & 6 (6)   & 8 (8)   & 0 (0)  & 1 (11)  \\
				\makebox[3cm][r]{5001-50000}    & 15 (7)   & 6 (6)   & 9 (8)    & 1 (17)   & 0 (0)   \\
				\makebox[3cm][r]{$> 50001$}    & 2 (1)      & 1 (1)   & 1 (1)     & 0 (0)  & 0 (0)   \\
				
				\rowcolor{Lavender} \multicolumn{6}{l}{Lifestyle} \\
				\makebox[3cm][l]{Smoking} & & &  &  & \\
				\makebox[3cm][r]{Yes} & 72 (36)& 49 (52)&       23 (22)  & 2 (33) & 3 (33) \\
				\makebox[3cm][r]{No} & 127 (64) & 45 (48)&  78 (78) & 4 (67) & 6 (67)\\
				
				\makebox[3cm][l]{Substance abuse} & & &  & &  \\
				\makebox[3cm][r]{Yes} & 24 (12) & 15 (16)&   9 (9)   & 0 (0)  & 1 (11) \\
				\makebox[3cm][r]{No}  & 175 (88)& 79 (84)& 96 (91) & 6 (100) & 8 (89) \\
				
				\makebox[3cm][l]{Education} & & & &  &  \\
				\makebox[3cm][r]{$<$High school} & 49 (25) & 23 (24) & 26 (25) & 0 (0)  & 1 (11) \\
				\makebox[3cm][r]{High school}                 &121 (61)  & 62 (66)& 59 (56) & 5 (83)  & 8 (89)\\
				\makebox[3cm][r]{$\geq$College}         & 29 (14)& 9 (10)      & 20 (19) & 1 (17) & 0 (0) \\
				\hline
				
		\end{tabular} }
	\end{centering}
	\caption{Demographic, clinical, and behavioral characteristics of participants at their first visits in the overall sample and in the four clusters. }
	\label{tab:covariates}
\end{table}

%
\subsection{Results}
We applied BAGEL to the dataset with the same hyperparameters as in the simulation study.  The least-square summary of clustering estimated four clusters with the number of participants in each cluster being 94, 105, 6, 9, respectively.  Table \ref{tab:covariates} summarizes demographic, clinical, and behavioral characteristics of participants  in the four clusters at their first visits, and Supplementary Table T3 reports the frequency of the ART drugs used in the four clusters. 
Since clusters 3 and 4 only had few participants, our subsequent analyses and clinical interpretation focused on clusters 1 and 2.

\begin{figure}[ht!]
	\begin{centering}
		\begin{tabular}{cc}
			\includegraphics[width=.5\textwidth]{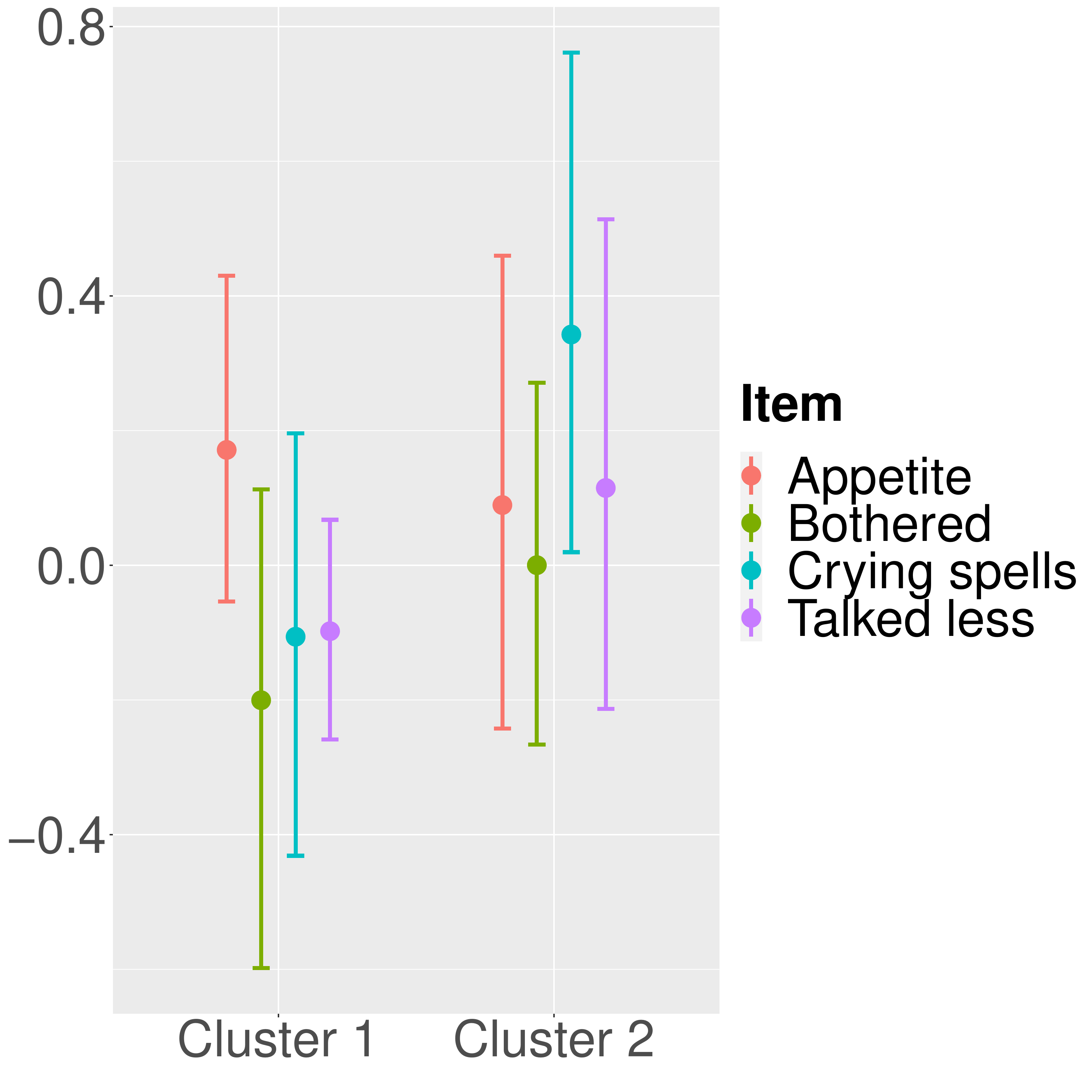}&\includegraphics[width=.5\textwidth]{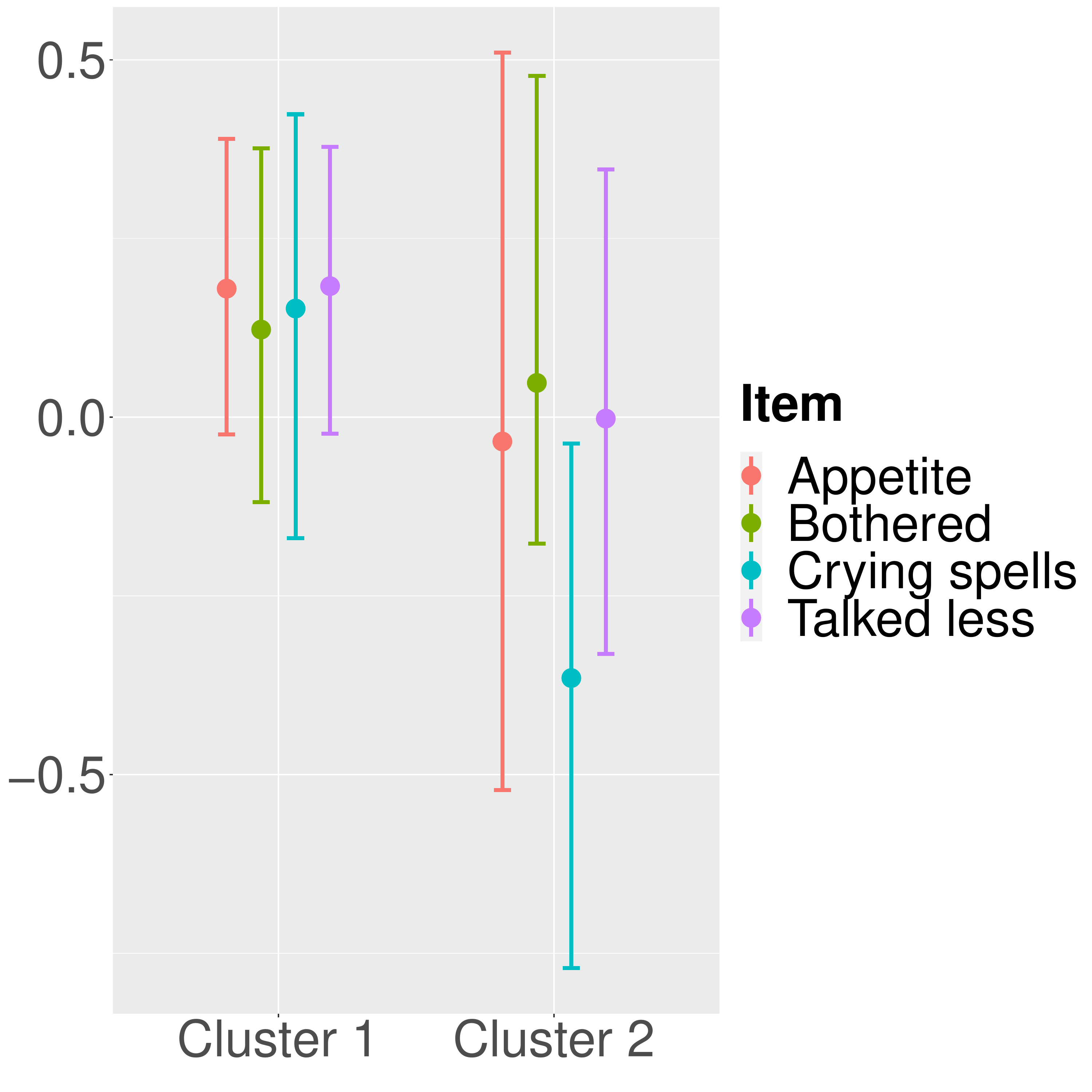}\\
			\hspace{-2cm} (a) Age & \hspace{-2cm} (b) CD4\\
			\includegraphics[width=.5\textwidth]{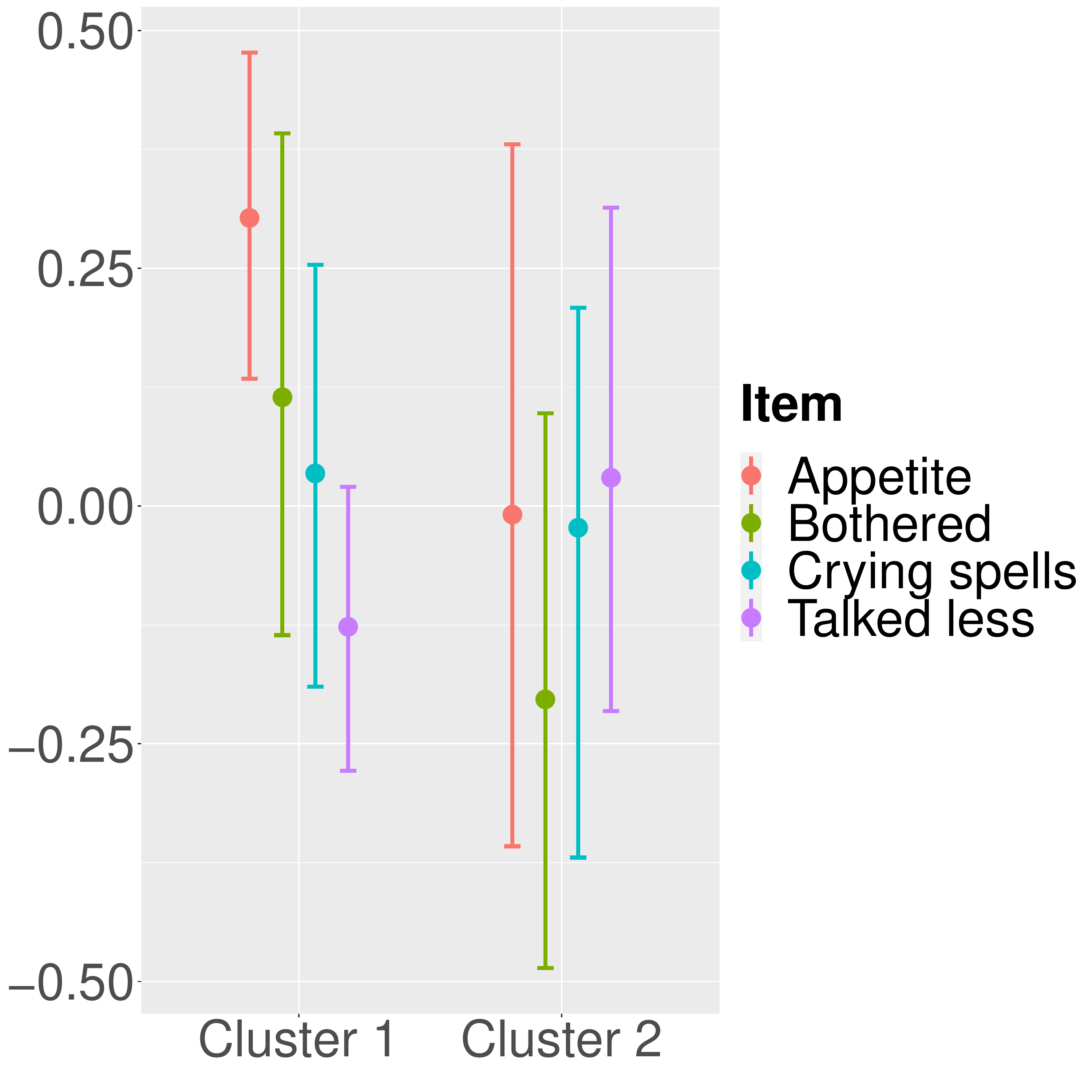}&\includegraphics[width=.5\textwidth]{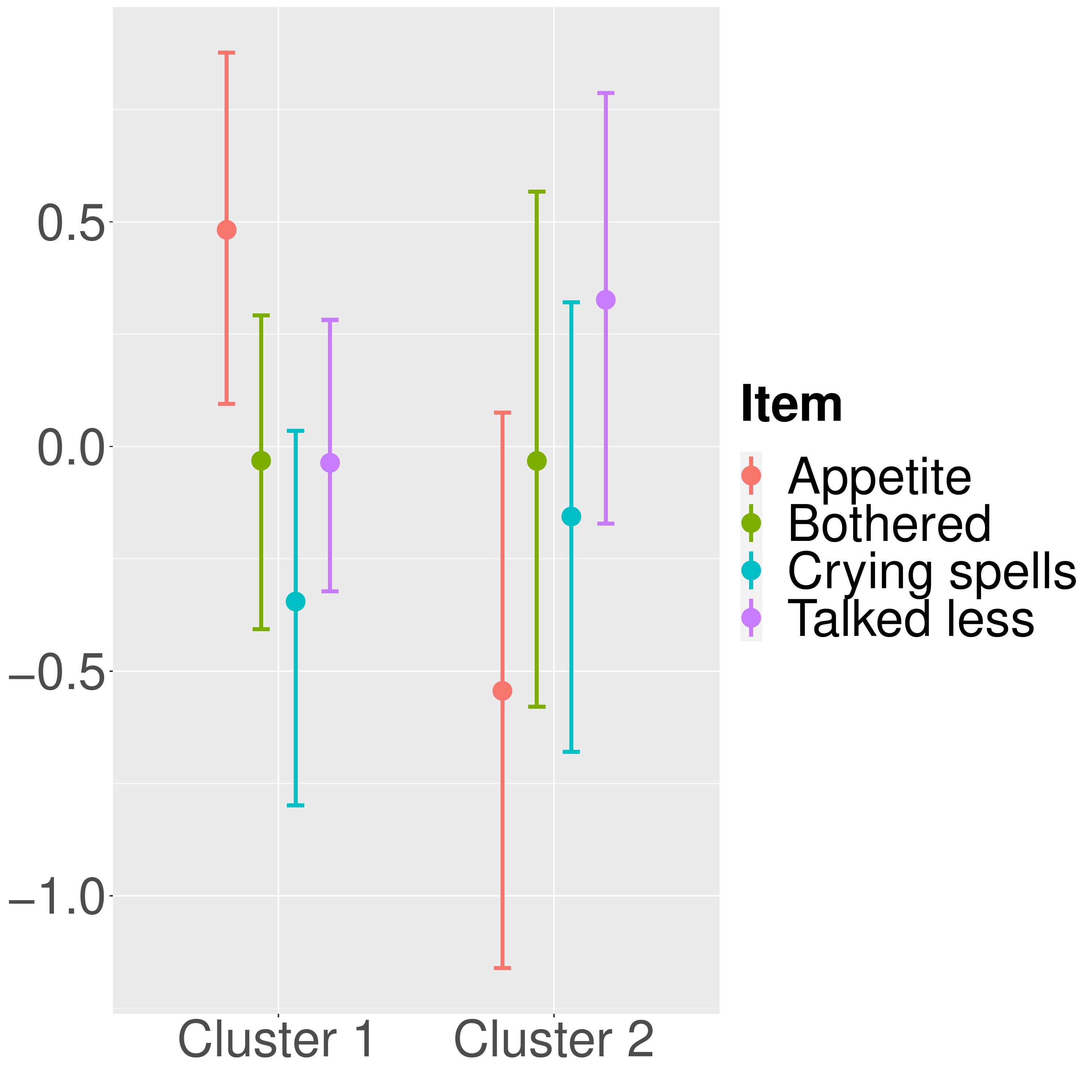}\\
			\hspace{-2cm} (c) VLOAD & \hspace{-2cm}  (d) Smoking
		\end{tabular}
	\end{centering}
	\caption{Posterior means and 95\% CIs for the estimated coefficients corresponding to age, CD4, viral load, and smoking. The dots represent the posterior means.}
	\label{fig:cov}
\end{figure}

Figure~\ref{fig:cov} plots posterior means and the corresponding 95\% CIs of the estimated effect sizes of age, CD4,  viral load, and smoking for four randomly selected depressive symptoms: ``appetite", ``bothered", ``crying spells", and ``talked less." As shown in Figure~\ref{fig:cov}, covariates have distinct effects in different clusters on different depression items. Panel (a) shows that younger people had less lack of appetite in both clusters 1 and 2, but 95\% CIs covered 0, indicating that the age effects were not significant on appetite in both clusters. In contrast, the effect of age on ``crying spells" was significant in cluster 2, but not in cluster 1, highlighting the heterogeneity among participants. Panel (b) 
indicates that a higher CD4 was associated with less ``crying spells" in cluster 2. Panels (c) and (d) show that a higher viral load and smoking were associated with more lack of appetite  in cluster 1, but not in cluster 2. As shown in Table \ref{tab:covariates}, the proportion of smoking participants in cluster 1 (52\%) is significantly higher than  that in cluster 2 (22\%), indicating that women who smoke are more likely to experience lack of appetite when viral load is high. These findings are consistent with the literature \citep{brink2010monitoring,taniguchi2014depression, clubreth2016associations,williams2020associations}.


\begin{figure}[ht!]
	\begin{centering}
		\begin{tabular}{cc}
			\includegraphics[width=.5\textwidth]{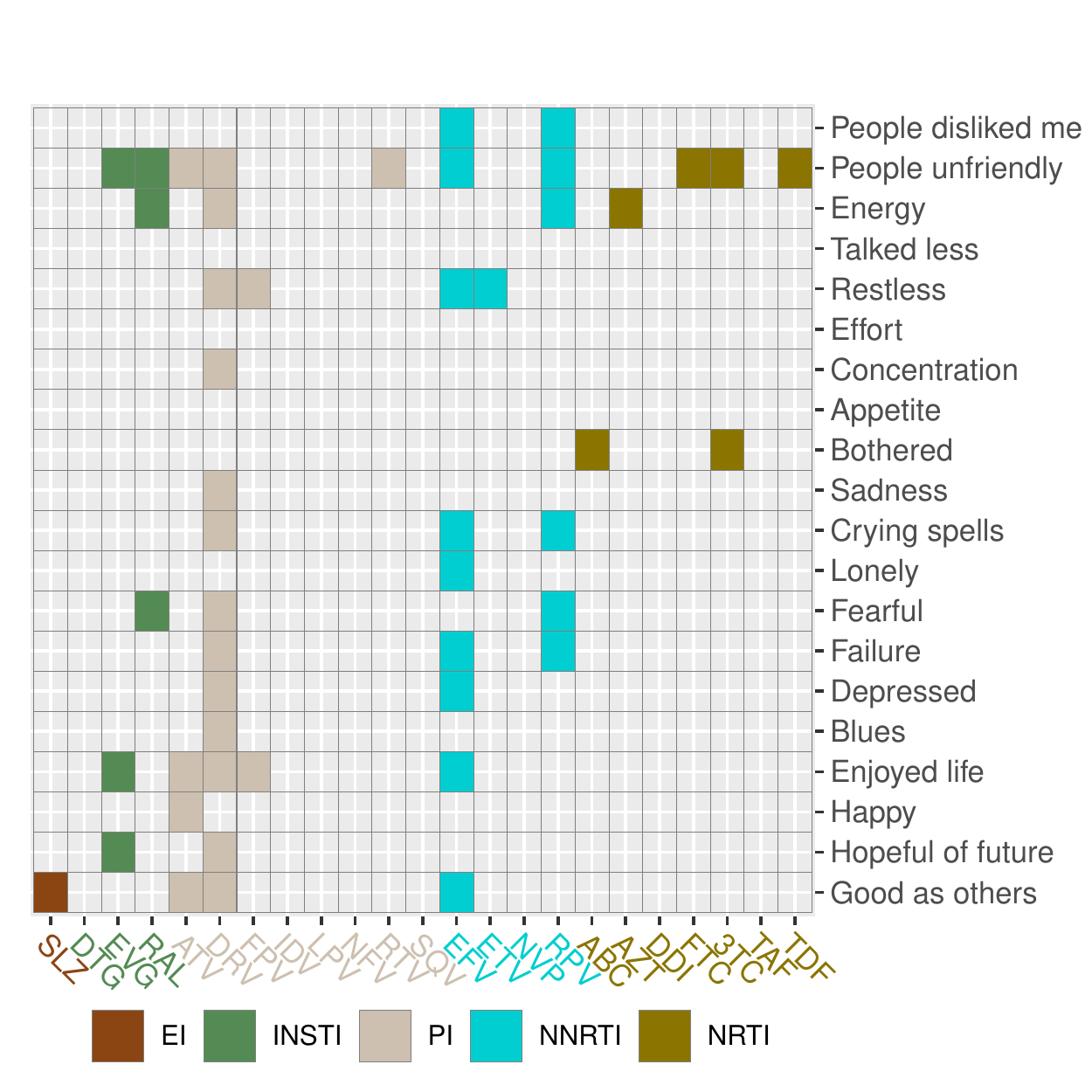}&\includegraphics[width=.5\textwidth]{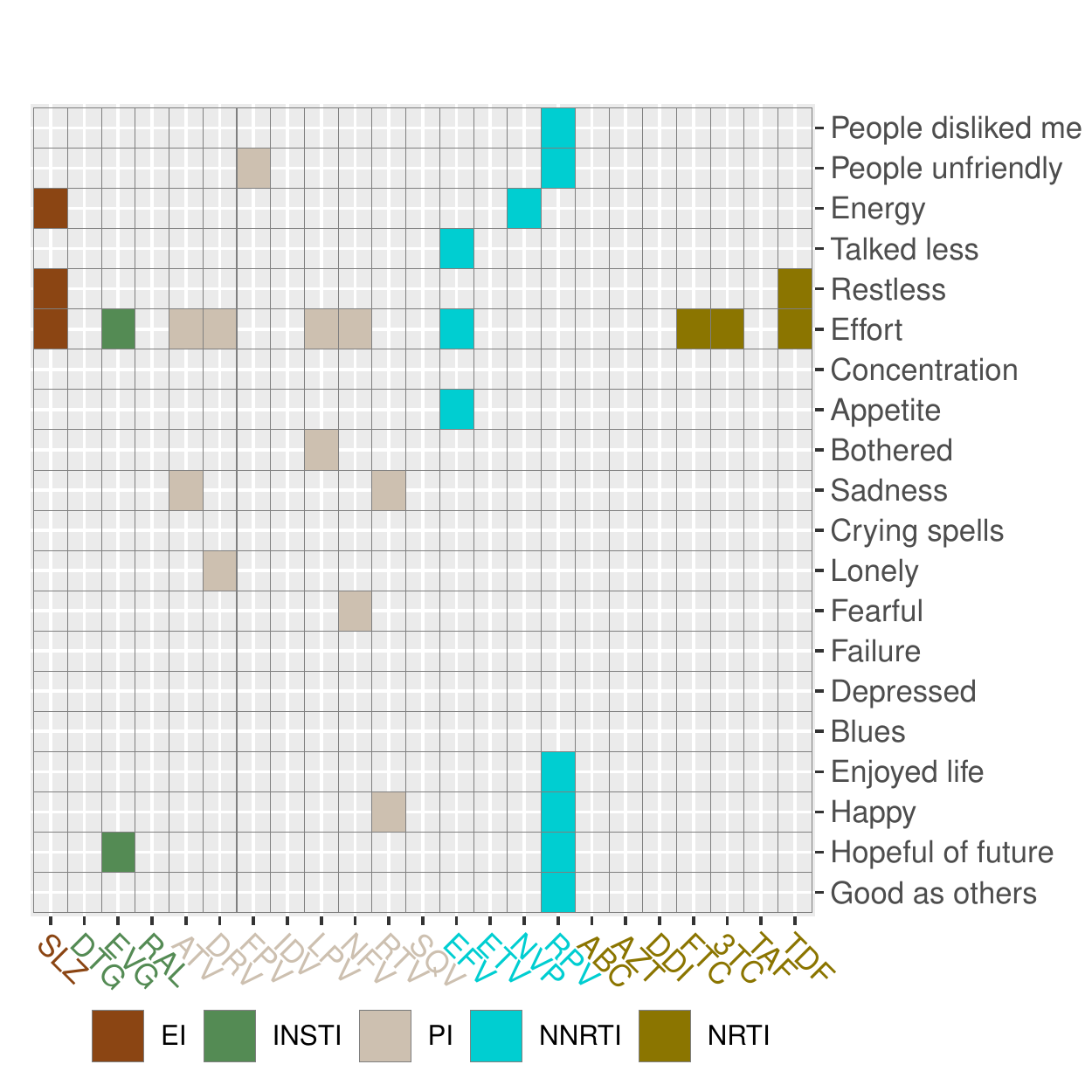}\\
			(a) Cluster 1 &(b) Cluster 2
		\end{tabular}
	\end{centering}
	\caption{Associations between ART drugs and depression items: a color-filled cell indicates that there exists significant association between the corresponding drug and depression item.}
	\label{fig:edge}
\end{figure}

Then we report the estimated $\tilde{R}_{h, dq}$, which represents if drug $d$ is significantly associated with depression item $q$ for participants in cluster $h$. Note that $\tilde{R}_{h, dq}$ only indicates whether the significant association exists. The sign and magnitude  are represented by $\bB_{ij}$ through $\bLambda_{ij}$, which will be summarized later. 
Figure~\ref{fig:edge} presents the associations between ART drugs and 20 depression items in clusters 1 and 2 corresponding to the least-square point estimate, showing that the drug-depression associations are different between clusters. Supplementary Figure F7 plots the posterior probabilities of $\{\tilde{R}_{h, dq}=1\}$ across all drugs and depression items. 
The most frequently used NRTI in the dataset, TDF, was associated with the symptom ``people unfriendly" in cluster 1, and ``restless" and ``effort" in cluster 2. 
Several clinical trials have reported the associations between TDF and depression \citep{squires2003tenofovir, mills2016switching}. 
The most frequently used NNRTI drug, EFV, was   associated with nearly half of the depression items in cluster 1 and three items in cluster 2. 
This is consistent with studies demonstrating associations  between EFV and depressive symptoms, such as 
suicidal behavior \citep{mollan2014association, bengtson2017relationship, arenas2018risk}. INSTI drugs have also been linked to psychiatric symptoms including depression and are among primary reasons for discontinuing INSTI treatment \citep{hoffmann2017higher, borghetti2018slc22a2, revuelta2018effectiveness}. For example,  \cite{harris2008exacerbation} reported several treatment-experienced HIV-seropositive   patients had significant exacerbation of pre-existing depression after starting to take RAL, an INSTI drug. The proposed BAGEL identified significant associations between two of the INSTI drugs (EVG and RAL) and several depression items in the two clusters. Compared to previous studies, BAGEL not only determines whether a certain drug is significantly associated with depression, but also which specific depression item that drug is associated with. Since two people with the same sum-score (total depression score) may have distinct symptoms that have different consequences and require different treatments, these more precise findings demonstrate the potential of BAGEL to facilitate precision medicine on treating HIV and its comorbidities including depression.

\begin{figure}[ht!]
	\begin{centering}
		\begin{tabular}{ccccc}
			\hspace{-.5cm}\includegraphics[width=.2\textwidth]{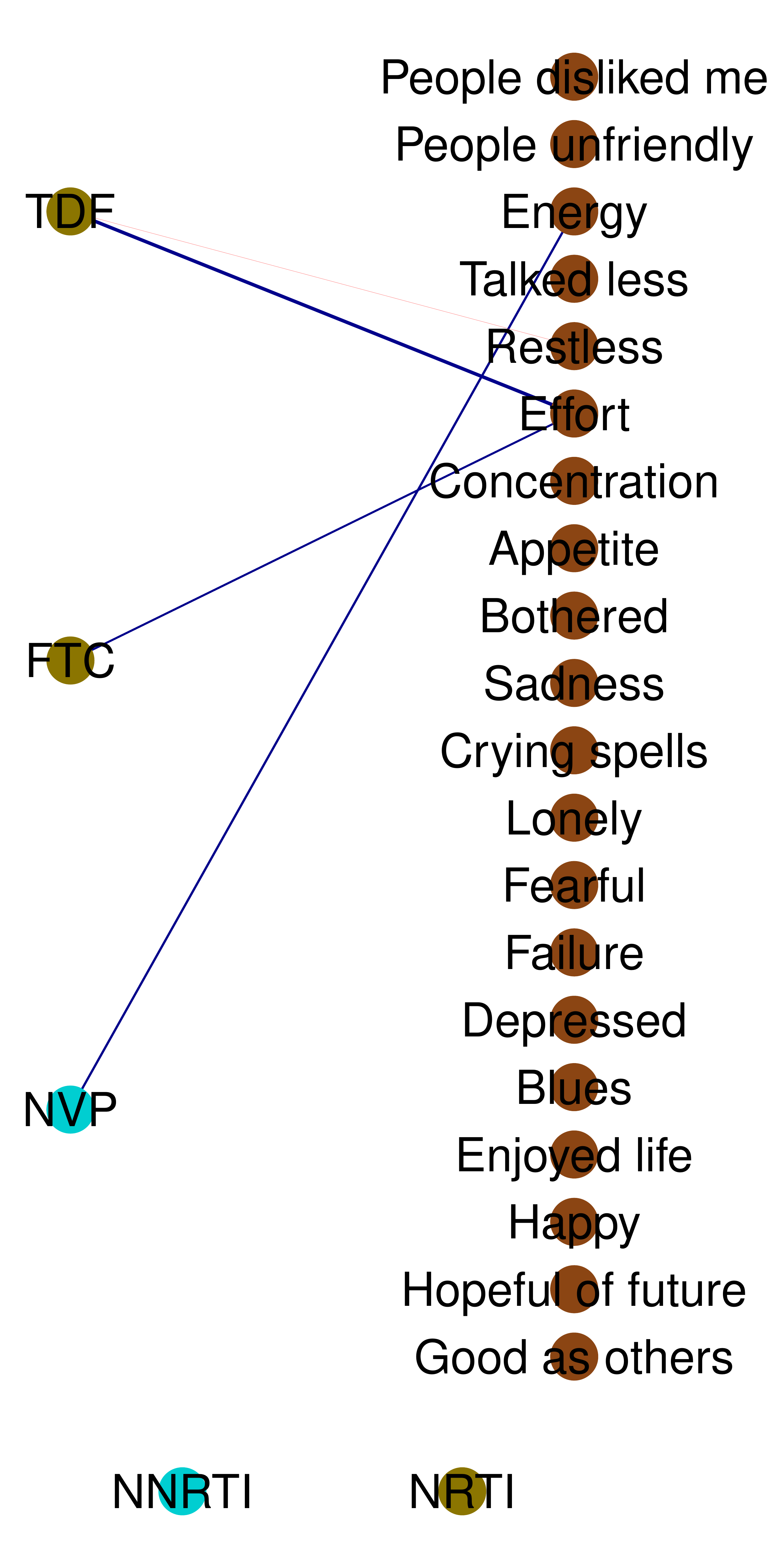}&\hspace{-.3cm} \includegraphics[width=.2\textwidth]{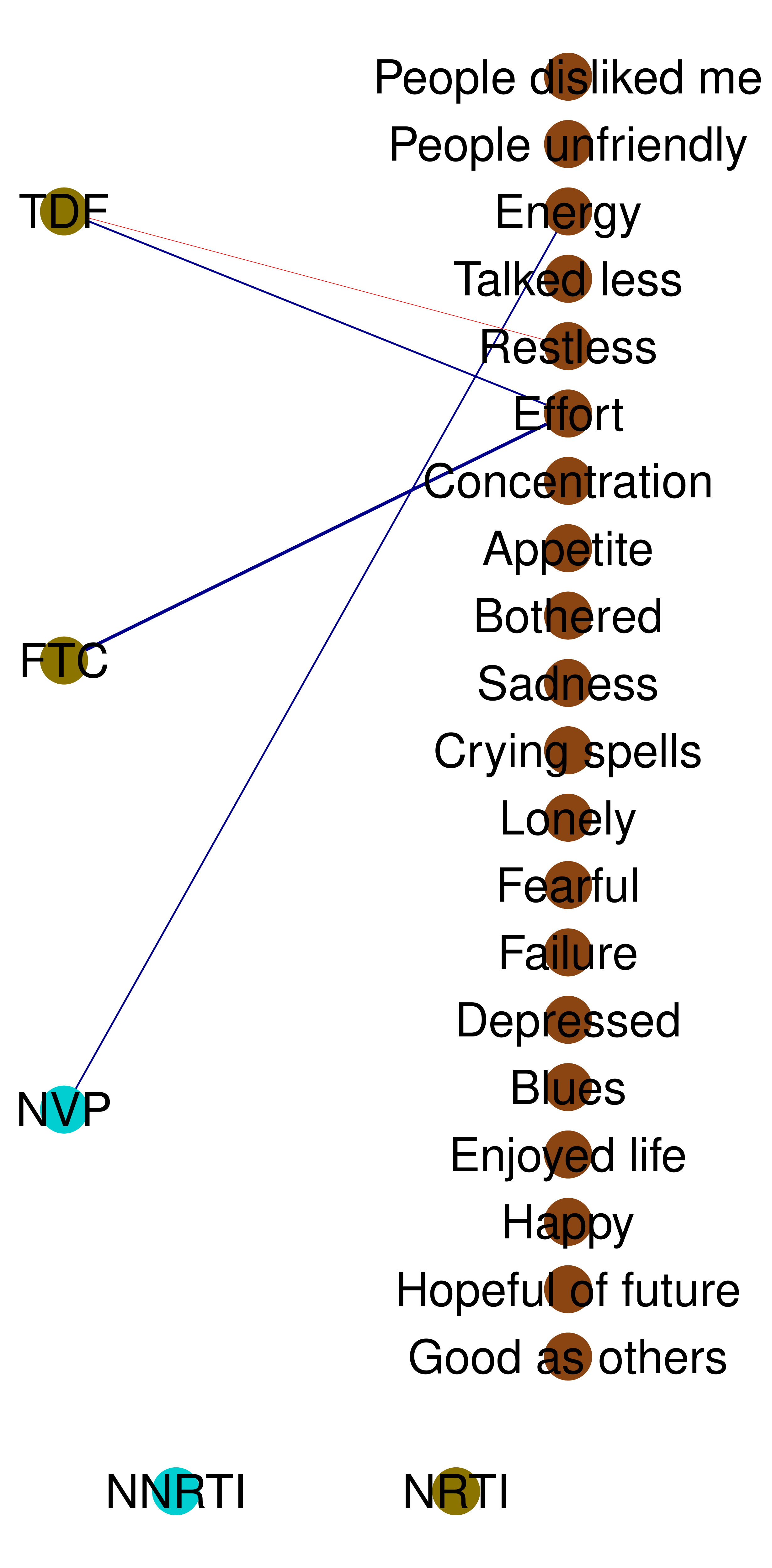}&\hspace{-.3cm}\includegraphics[width=.2\textwidth]{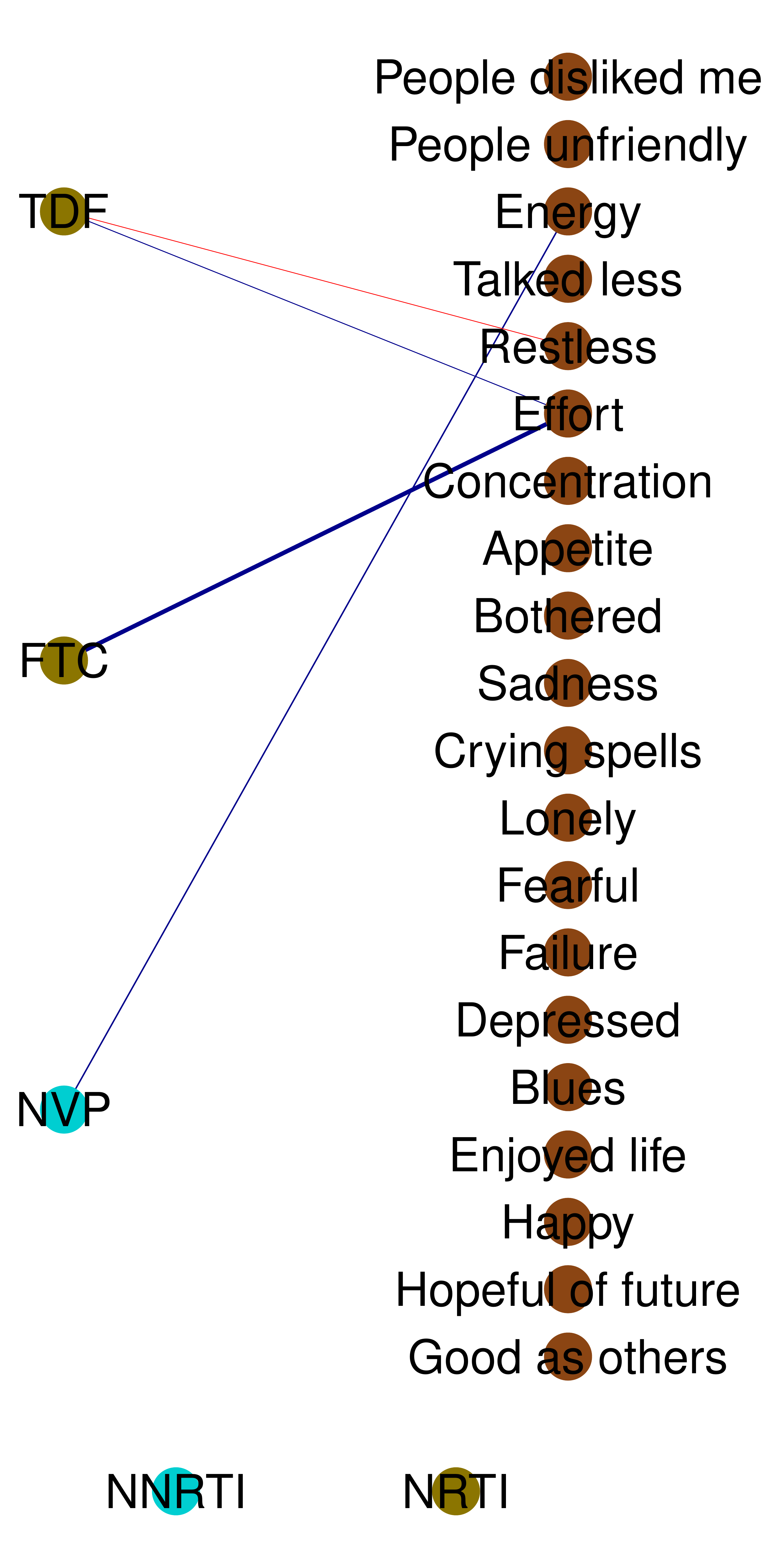}&\hspace{-.3cm}\includegraphics[width=.2\textwidth]{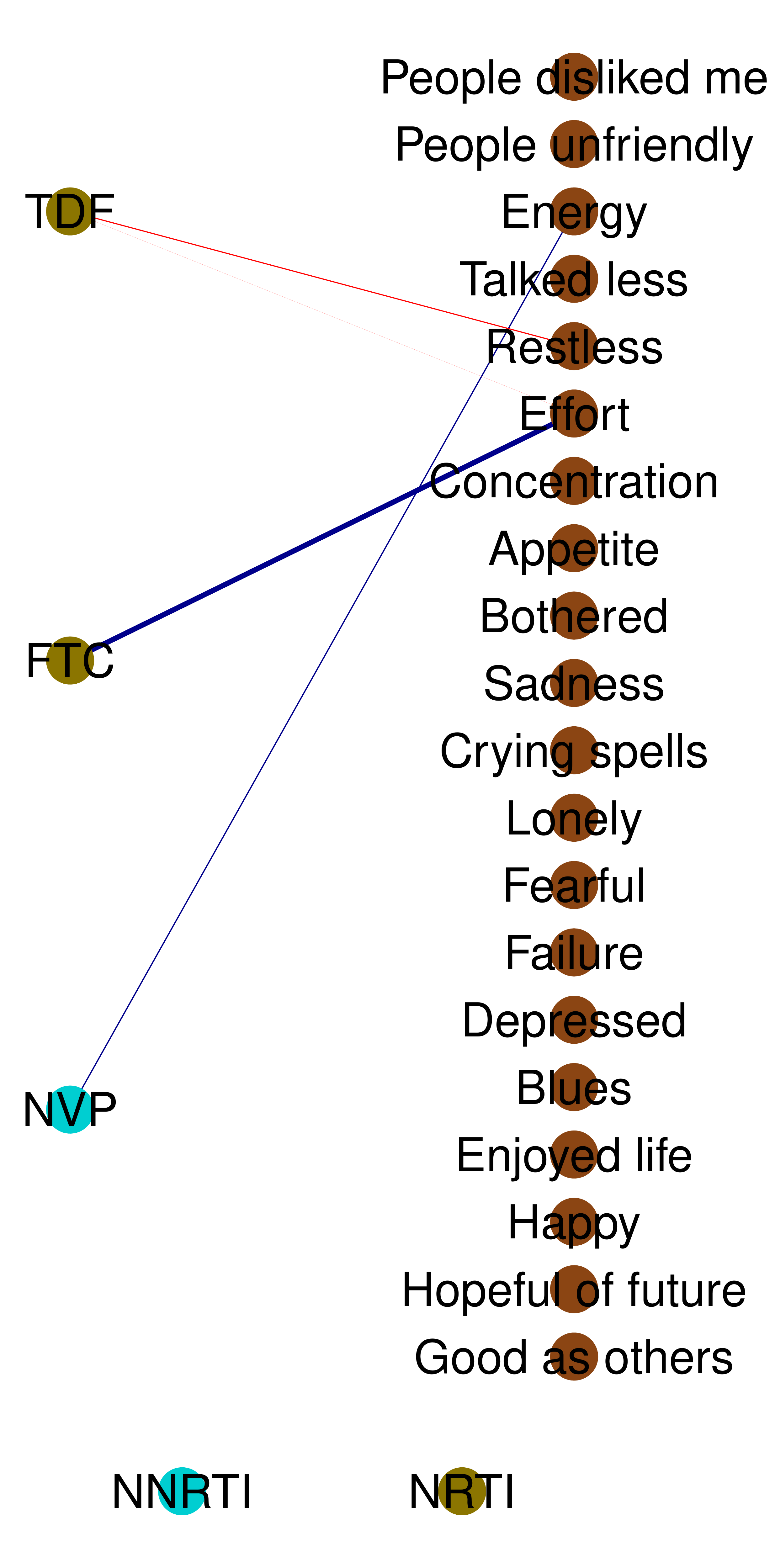}&\hspace{-.3cm}\includegraphics[width=.2\textwidth]{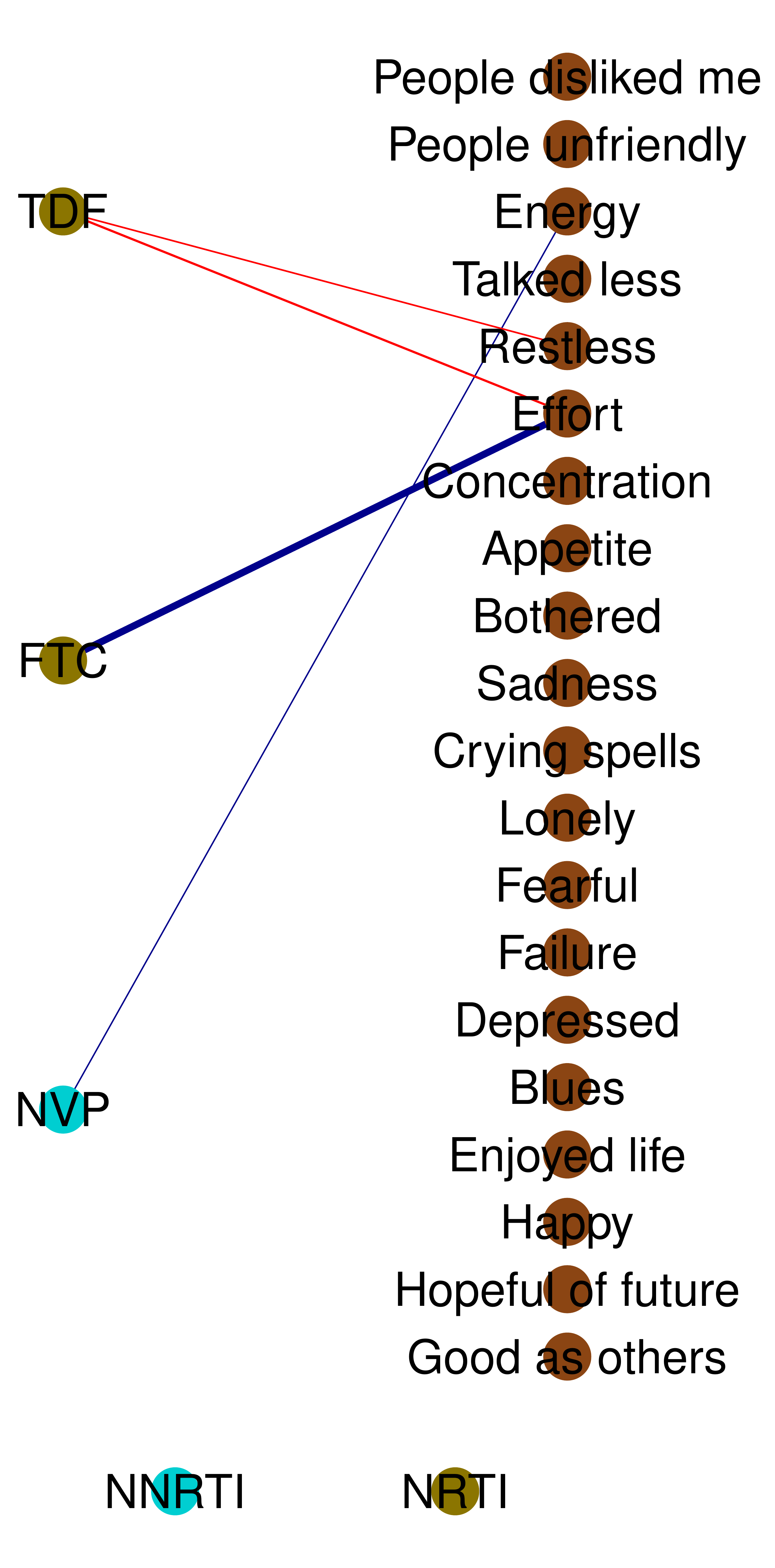}\\
			\hspace{-1cm}(a) Visit 1 &(b) Visit 2 & (c) Visit 3 & (d) Visit 4 & (e) Visit 5 \\
		\end{tabular}
	\end{centering}
	\caption{Drug effect on depression symptoms for participant 1. Blue lines indicate that the ART drug is associated with less symptomatology and red lines indicate that the ART drug is associated with more symptomatology. The width of the line indicates the magnitude of the association. }
	\label{fig:visit}
\end{figure}

Next we report the estimation of drug effects $\bB_{ij}$ on depressive symptoms for participant $i$ at visit $j$. 
For illustration, we randomly selected one participant from each of clusters 1 and 2, called participant 1 and participant 2 hereafter.   Figure \ref{fig:visit}  depicts the
associations between ART drugs and item-level depressive symptoms over time for participant 1. 
Blue lines indicate that the ART drug is associated with less symptomatology, indicating favorable effects in reducing depressive symptoms, and red lines indicate that the ART drug is associated with more symptomatology, indicating unfavorable effects on symptoms.  The width of the line is proportional to the magnitude of the association. The drug effects on depressive symptoms change over time, in both magnitude and sign. Particularly, participant 1 had 5 visits and used two NRTIs (TDF+FTC) and one NNRTI (NVP) in all 5 visits. TDF was associated with more symptomatology on ``restless", and the negative effect was magnified over time: 0.055 at the first visit, then 0.211, 0.309, 0.47 at visits 2-4, until 0.693 at the last visit, indicating that the long-term use of TDF could cause lasting and worse sleeping issue for people with HIV.  TDF was associated with less symptomatology on ``effort" at the first visit with the corresponding $B_{ij, dq}$ being -1.472, but the beneficial effect kept decreasing at the second and third visits with effects being -0.795 and -0.379, respectively. At the fourth visit, the beneficial effect of  TDF on ``effort" became harmful with the effect being 0.013, then became worse at the fifth visit with the effect being 0.941. In contrast, FTC had beneficial effect on ``effort", and the effect kept increasing over time. NVP was also beneficial for ``energy" in all visits. In general, failure to suppress viral load and severe side effects are two major reasons for treatment switch. This participant's viral loads were undetectable in all five visits and she was consistently reported use of  TDF+FTC+NVP, which agreed with our finding that TDF+FTC+NVP was well tolerated in terms of depressive symptoms. However, if sleeping problem was the major concern for this woman, and such symptom might not be reflected by the CESD sum-score since the overall score could be dominated by positive effects of drugs on other  symptoms, then BAGEL could guide the physician to switch treatment by considering another NRTI to replace TDF.

\begin{figure}[ht!]
	\begin{centering}
		\begin{tabular}{ccc}
			\includegraphics[width=.33\textwidth]{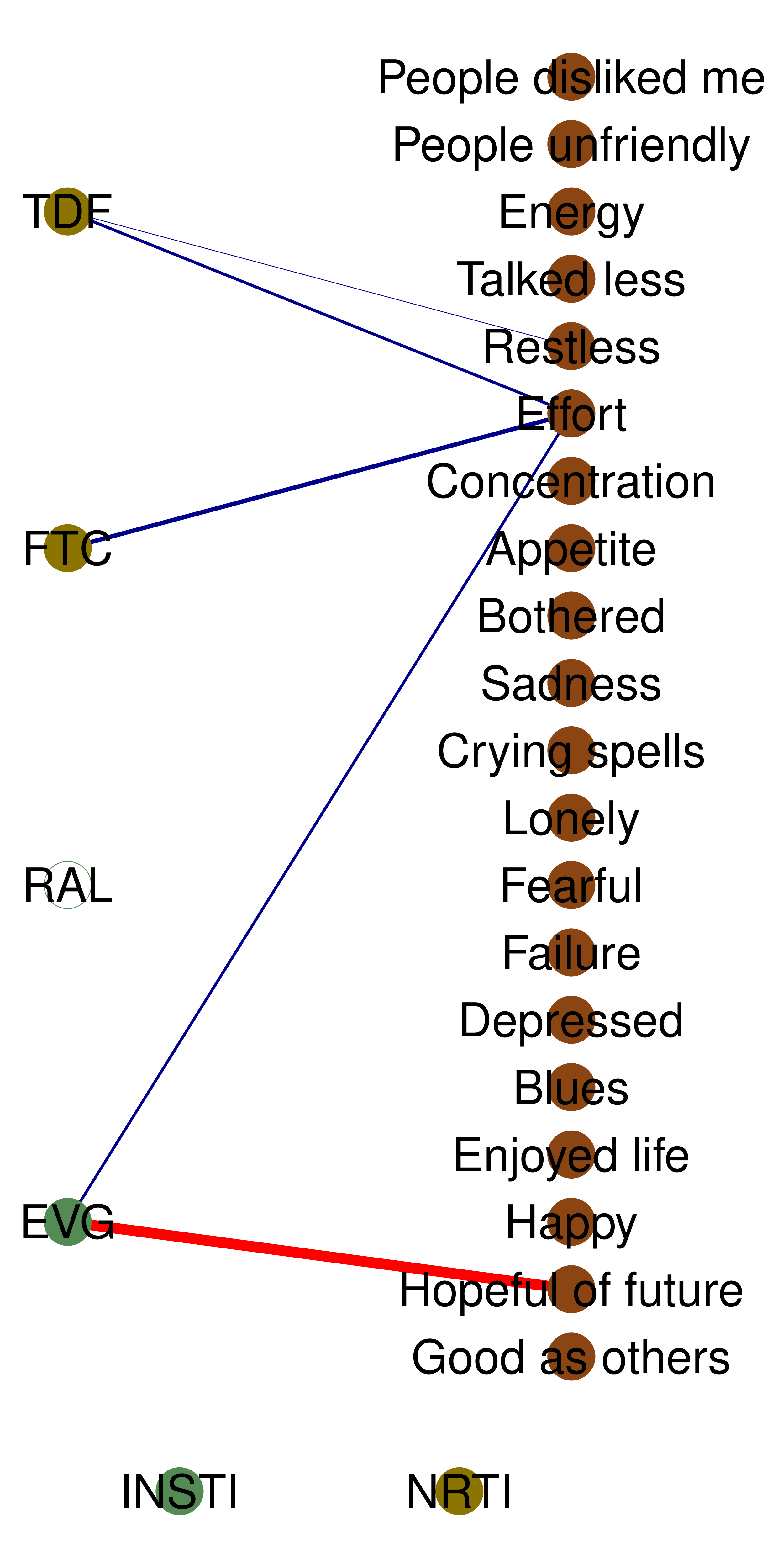}&\includegraphics[width=.33\textwidth]{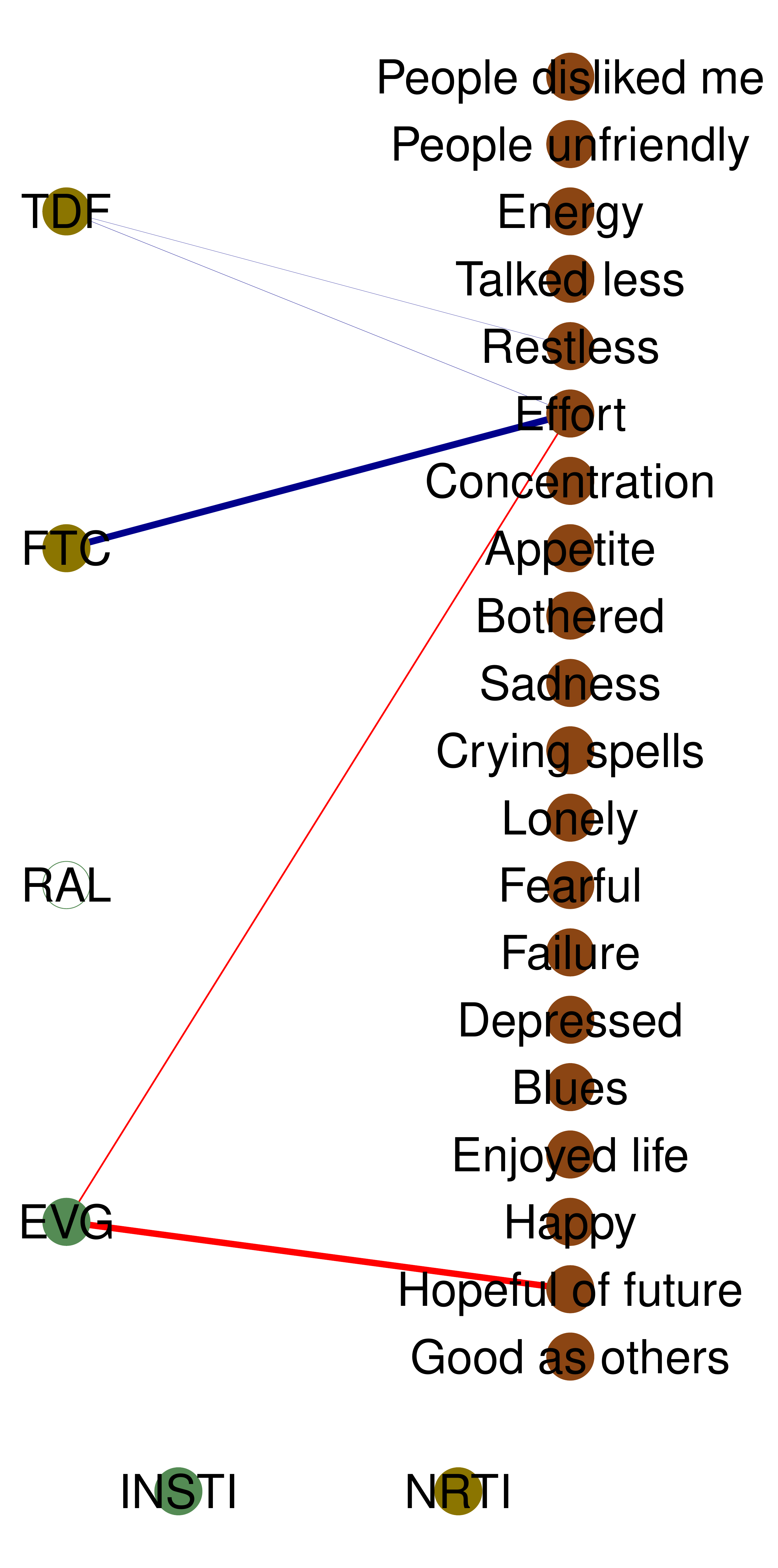}&\includegraphics[width=.33\textwidth]{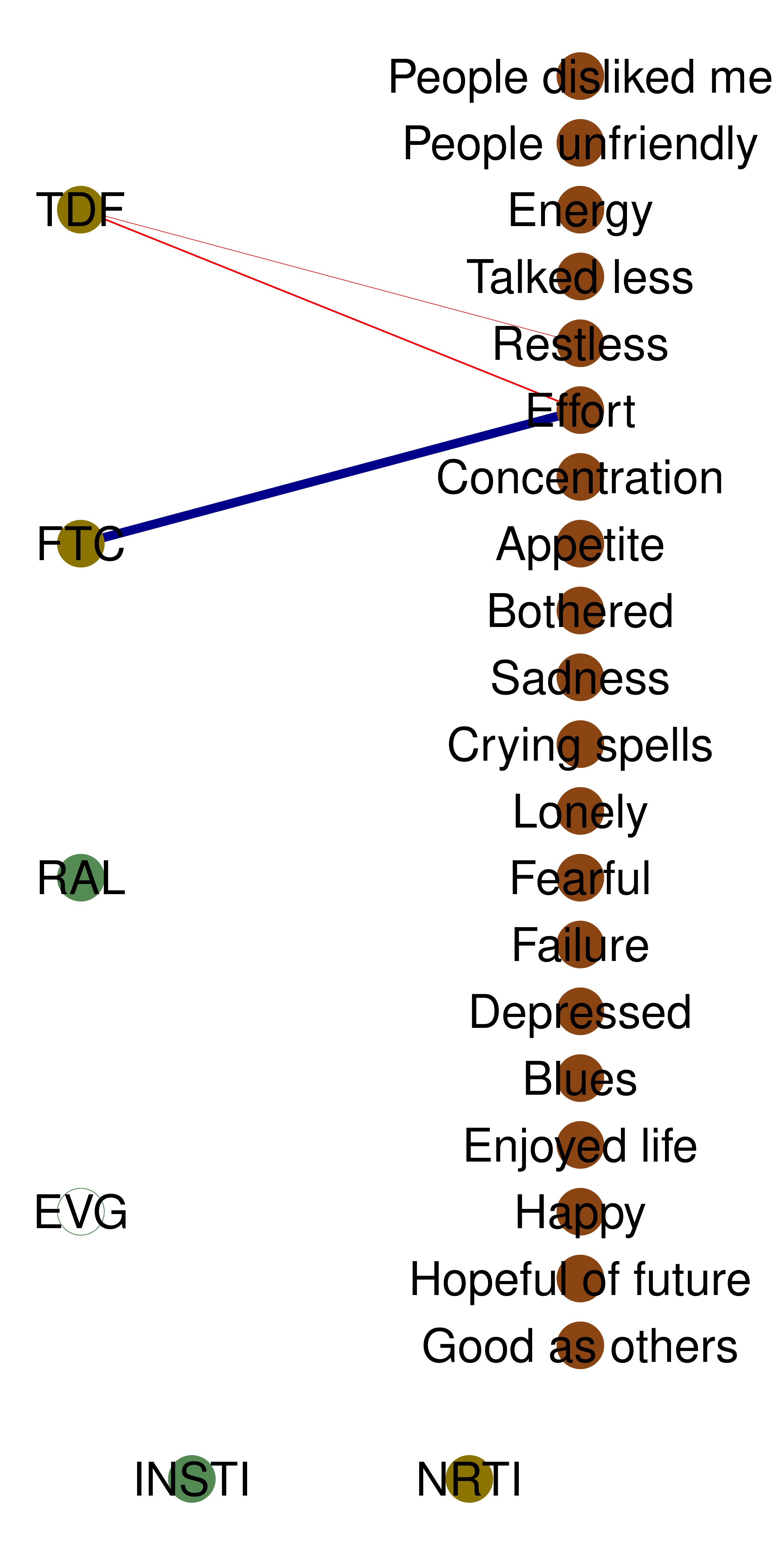}\\
			(a) Visit 1 &(b) Visit 2 & (c) Visit 3 \\
		\end{tabular}
	\end{centering}
	\caption{Drug effect on depression symptoms for participant 2. Blue lines indicate that the ART drug is associated with less symptomatology and red lines indicate that the ART drug is associated with more symptomatology. The width of the line indicates the magnitude of the association. }
	\label{fig:visit2}
\end{figure}

Figure \ref{fig:visit2} plots the associations between ART drugs and item-level depressive symptoms over time for  participant 2. This participant had three visits: at the first two visits, she reported using two NRTIs (TDF and FTC) and one INSTI (EVG); at her last visit, she reported using the same two NRTIs but switched to another INSTI (RAL). Since RAL is not significantly associated with any depressive symptom in cluster 2, there is no edge connecting RAL with symptoms. TDF was associated with less symptomatology in two somatic symptoms (``restless" and ``effort") at her first visit, then the beneficial effects decreased at her second visit. As she continued to take TDF, the effects of TDF turned negative on these two depressive symptoms at her last visit. TDF was reported to have limited neurotoxic effects due to its low central nervous system penetrance \citep{best2012low}. However, our finding highlights the need of evaluating the long-term effect of TDF on depressive symptoms in people with HIV, especially somatic symptoms. EVG had a positive effect on ``effort" and a negative effect on  ``hopeful of future" at her first visit, then the effect on ``effort" became negative at her second visit. Previous clinical trials  \citep{cohen2011randomized,abers2014neurological} reported psychiatric adverse events for participants treated with EVG. 
As shown in Figure \ref{fig:visit2}, this participant switched from EVG to RAL, indicating that those negative effects of EVG on depression may be the major reason for this switch because her viral loads were well suppressed across all visits and hence unlikely caused the switch.

\begin{figure}[ht!]
	\begin{centering}
		\includegraphics{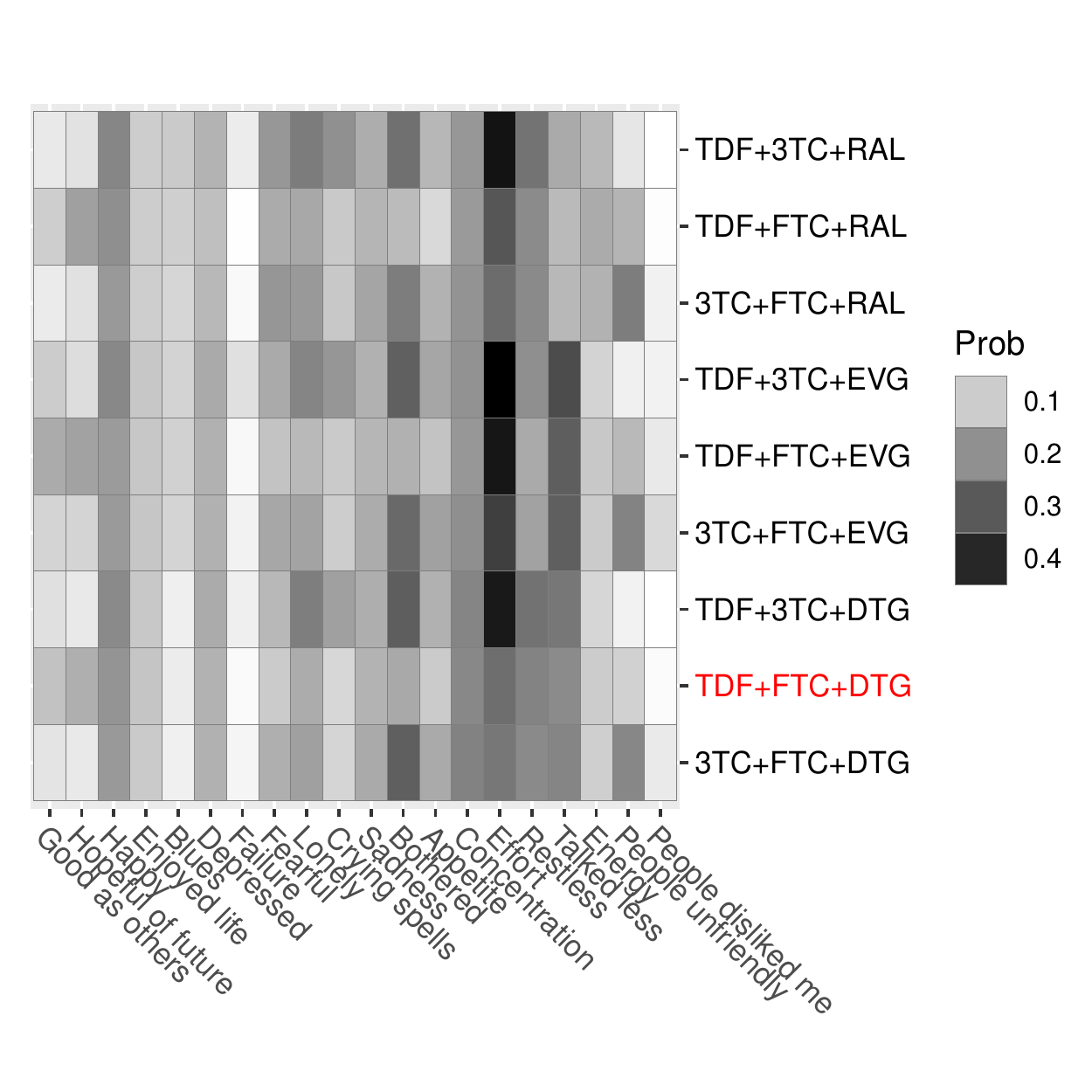}\\
		\vspace{-1cm}
	\end{centering}
	\caption{Posterior predictive probabilities of depression items at participant 2's third visit given the data from her first two visits and various combinations of ART drugs. TDF+TFC+DTG is the optimal drug combination based on the utility score since it leads to the smallest $\Delta(\bz)$. }
	\label{fig:pred}
\end{figure}

Lastly, we demonstrate how BAGEL can  guide more informed and effective individualized ART drug selection for better depression control in HIV clinical practice.  Assume that we have observed data for  participant $i$ at her first $J_i$ visits, denoted by $\mathcal{D}_i=\{U_{ijq}, \bZ_{ij}\}_{q=1, j=1}^{Q, J_i}$, and we aim to find an optimal drug combination at her $(J_i+1)$-st visit.  For example, pretending that participant 2 has data reported up to her second visit,  i.e., $J_i=2$, the analysis 
with BAGEL can assist physician in drug management at her third visit.  The US department of Health and Human Service has recommended two NRTI drugs as backbone with an additional INSTI drug as   first-line therapy (https://aidsinfo.nih.gov/guidelines). 
Assume that at her third visit, the physician wants to choose two NRTIs from TDF, 3TC, and FTC, and choose one INSTI from RAL, EVG, and DTG, resulting in a total of 9 possible combinations.  Denote the  posterior predictive probability of the $q$-th depression score at her third visit being larger than 0 under drug combination $\bz$ by $\pi_{iq}(\bz)=\mathrm{Pr}(U_{i3q}>0 \mid \bZ_{i3}=\bz, \mathcal{D}_i)$, $q=1, \dots, Q$ which is shown in Figure \ref{fig:pred}.
As an example, we define an individualized utility score $\Delta_i(\bz)$ as the criteria of choosing an optimal drug combination based on the sum of  posterior predictive probabilities for all depression items $\Delta_i(\bz)=\sum_q \pi_{iq}(\bz)$. Clearly, a drug combination is less desirable if it has a higher score. For simplicity, we first compare two candidate drug combinations  that she took at her first two visits (Figure \ref{fig:visit2}), $\bz_1$=TDF+FTC+EVG and $\bz_2$=TDF+FTC+RAL. We found 
$\Delta_i(\bz_1)=2.988$ with the top three adverse effects being ``effort" (0.439), ``talked less" (0.290), and ``concentration" (0.188) whereas
$\Delta_i(\bz_2)=2.714$ with the top three adverse effects being ``effort" (0.307), ``restless" (0.210), and ``happy" (0.202). Since $\bz_2$=TDF+FTC+RAL leads to a smaller utility score, we claim that $\bz_2$=TDF+FTC+RAL is better than $\bz_1$=TDF+FTC+EVG in controlling overall depressive symptoms for this participant and hence would recommend a drug switch from EVG to RAL at her third visit. As shown in Figure \ref{fig:visit2}, the physician's actual choice was indeed RAL, which agreed with our prediction. However, if DTG was considered in the comparison, we would recommend the
drug combination TDF+TFC+DTG since it is expected to cause the least depressive symptomatology: $\Delta_i(\bz)=2.646$ with the top three adverse effects being ``effort" (0.261), ``restless" (0.224), and ``concentration" (0.215). 


Note that in our simple illustrative example, the defined individualized utility score $\Delta_i(\bz)$ simply sums over all depression items with equal weights. In clinical practice, different categories of depression symptoms might impact clinical utility differently. For example, somatic symptoms including ``energy", ``talked less", ``restless", ``effort", ``concentration", ``appetite", and ``bothered",  
have been shown more likely to cause discontinuation of treatment for people with HIV \citep{kapfhammer2006somatic,taibi2013sleep}. Therefore, 
the utility score $\Delta_i(\bz)$ can be adjusted to impose more weights on somatic symptoms chosen according to the physician's opinion or individual preference of people with HIV.  
For example, let $\Xi$ denote the set of somatic symptoms, the utility score can be alternatively defined as $\widetilde{\Delta}_i(\bz)=\sum_{q\in \Xi} \pi_{iq}(\bz)$ if the physician wants to focus on somatic symptoms.  Again, take participant 2 for illustration.  We computed $\widetilde{\Delta}_i(\bz)$ for all 9 drug combinations and concluded that TDF+FTC+DTG was  still the optimal choice.  Finally, we remark that the flexibility of choosing utility score is enabled by one of the prominent features of BAGEL, i.e., precise probabilistic characterization of ART drug effects on item-level depressive symptoms.


\section{Conclusion}
\label{sec:discuss}
To better understand the long-term effects of ART or ART drug switches on item-level depressive symptoms longitudinally and facilitate HIV precision medicine, we developed BAGEL, a novel Bayesian graphical model that  estimates   longitudinal drug effects on depression while accounting for participants' heterogeneity as well as demographic, clinical, and behavior characteristics. 
Through simulation studies and analysis of the WIHS dataset, we have demonstrated that BAGEL accurately estimates the longitudinal drug  effects, yields meaningful and interpretable results, and has the potential to assist physicians' decisions on personalized ART drug prescriptions. In addition, we have made the code that implements BAGEL publicly available so that users can apply BAGEL to datasets in a similar setup. 

We focus on the effects of ART drugs on depressive symptomatology in this paper, but BAGEL is also helpful to understand other ART-related complications such as cognitive impairment  using the model of $\bY_{ij}$ directly since cognition outcomes are continuous. BAGEL can also be easily extended to incorporate genetic polymorphisms, an important factor in studies related to mental health and HIV as the use of certain ART drugs in the setting of specific genetic polymorphisms can increase risk for adverse effects on mental health. For example, EFV combined with polymorphisms in CYP2B6 and CYP2A6 increases the risk of suicidality \citep{bengtson2017relationship, mollan2017race}. DTG combined with polymorphisms in SLC22A2 is associated with psychiatric symptoms \citep{borghetti2018slc22a2}. 
In the proposed model, polymorphisms can be incorporated as time-invariant covariates, i.e., being part of $\bX_{ij}$. With this setup, we inlcude the main effects of polymorphisms (via $\Xb_{ij}\bbeta_i$) as well as their interactions with ART drugs (via $\bZ_{ij}\bB_{ij}$).

BAGEL is based on a longitudinal directed acyclic graphical model. Although specifically motivated by the WIHS application,  it can be applied to other biomedical longitudinal studies or to other fields.
For example, in sports medicine, individual monitoring of
stress and recovery provides useful information to
prevent injuries and illnesses in athletes, and a number of longitudinal studies have been conducted in sports such as soccer and basketball \citep{jones2017training}. BAGEL can be applied to such longitudinal datasets to study the impact of psychosocial stress on the risk of sports injuries adjusting for physical stress effect.

\section*{Acknowledgment}

This work was supported by the Johns Hopkins University Center for AIDS Research NIH/NIAID fund (P30AI094189) 2019 faculty development award to Dr. Xu, NSF 1940107 to Dr. Xu, NSF DMS1918854 to Drs. Xu and Rubin, and NSF DMS1918851 to Dr. Ni. Data in this manuscript were collected by the Women’s Interagency HIV Study, now the MACS/WIHS Combined Cohort Study (MWCCS). The contents of this publication are solely the responsibility of the authors and do not represent the official views of the National Institutes of Health (NIH). MWCCS (Principal Investigators): Data Analysis and Coordination Center (Gypsyamber D’Souza, Stephen Gange and Elizabeth Golub), U01-HL146193; Metropolitan Washington CRS (Seble Kassaye and Daniel Merenstein), U01-HL146205. The MWCCS
is funded primarily by the National Heart, Lung, and Blood Institute (NHLBI), with additional co-funding from the Eunice Kennedy Shriver National Institute Of Child Health \& Human Development (NICHD), National Human Genome Research Institute (NHGRI), National Institute On Aging (NIA), National Institute Of Dental \& Craniofacial Research (NIDCR), National Institute Of Allergy And Infectious Diseases (NIAID), National Insti- tute Of Neurological Disorders And Stroke (NINDS), National Institute Of Mental Health (NIMH), National Institute On Drug Abuse (NIDA), National Institute Of Nursing Research (NINR), National Cancer Institute (NCI), National Institute on Alcohol Abuse and Alcoholism (NIAAA), National Institute on Deafness and Other Communication Disorders (NIDCD), National Institute of Diabetes and Digestive and Kidney Diseases (NIDDK).

\bibliographystyle{apalike}
\bibliography{graphic}

\newcommand{\noop}[1]{}
\begin{thebibliography}{}

\bibitem[Abers et~al., 2014]{abers2014neurological}
Abers, M.~S., Shandera, W.~X., and Kass, J.~S. (2014).
\newblock Neurological and psychiatric adverse effects of antiretroviral drugs.
\newblock {\em CNS drugs}, 28(2):131--145.

\bibitem[Adimora et~al., 2018]{adimora2018cohort}
Adimora, A.~A., Ramirez, C., Benning, L., Greenblatt, R.~M., Kempf, M.-C.,
  Tien, P.~C., Kassaye, S.~G., Anastos, K., Cohen, M., Minkoff, H., et~al.
  (2018).
\newblock Cohort profile: the women’s interagency hiv study (wihs).
\newblock {\em International journal of epidemiology}, 47(2):393--394i.

\bibitem[Albert and Chib, 1993]{albert1993bayesian}
Albert, J.~H. and Chib, S. (1993).
\newblock Bayesian analysis of binary and polychotomous response data.
\newblock {\em Journal of the American statistical Association},
  88(422):669--679.

\bibitem[Arenas-Pinto et~al., 2018]{arenas2018risk}
Arenas-Pinto, A., Grund, B., Sharma, S., Martinez, E., Cummins, N., Fox, J.,
  Klingman, K.~L., Sedlacek, D., Collins, S., Flynn, P.~M., et~al. (2018).
\newblock Risk of suicidal behavior with use of efavirenz: results from the
  strategic timing of antiretroviral treatment trial.
\newblock {\em Clinical Infectious Diseases}, 67(3):420--429.

\bibitem[Bacon et~al., 2005]{bacon2005women}
Bacon, M.~C., Von~Wyl, V., Alden, C., Sharp, G., Robison, E., Hessol, N.,
  Gange, S., Barranday, Y., Holman, S., Weber, K., and Young, M.~A. (2005).
\newblock The women's interagency hiv study: an observational cohort brings
  clinical sciences to the bench.
\newblock {\em Clinical and diagnostic laboratory immunology},
  12(9):1013--1019.

\bibitem[Barkan et~al., 1998]{barkan1998women}
Barkan, S.~E., Melnick, S.~L., Preston-Martin, S., Weber, K., Kalish, L.~A.,
  Miotti, P., Young, M., Greenblatt, R., Sacks, H., and Feldman, J. (1998).
\newblock The women's interagency hiv study.
\newblock {\em Epidemiology}, pages 117--125.

\bibitem[Bengtson et~al., 2016]{bengtson2016disparities}
Bengtson, A.~M., Pence, B.~W., Crane, H.~M., Christopoulos, K., Fredericksen,
  R.~J., Gaynes, B.~N., Heine, A., Mathews, W.~C., Moore, R., Napravnik, S.,
  Safren, S., and Mugavero, M.~J. (2016).
\newblock Disparities in depressive symptoms and antidepressant treatment by
  gender and race/ethnicity among people living with hiv in the united states.
\newblock {\em PloS one}, 11(8):e0160738.

\bibitem[Bengtson et~al., 2017]{bengtson2017relationship}
Bengtson, A.~M., Pence, B.~W., Mollan, K.~R., Edwards, J.~K., Moore, R.~D.,
  O?CLEIRIGH, C., Eaton, E.~F., Eron, J.~J., Kitahata, M.~M., Mathews, W.~C.,
  et~al. (2017).
\newblock The relationship between efavirenz as initial antiretroviral therapy
  and suicidal thoughts among hiv-infected adults in routine care.
\newblock {\em Journal of acquired immune deficiency syndromes (1999)},
  76(4):402.

\bibitem[Best et~al., 2012]{best2012low}
Best, B.~M., Letendre, S.~L., Koopmans, P., Rossi, S.~S., Clifford, D.~B.,
  Collier, A.~C., Gelman, B.~B., Marra, C.~M., McArthur, J.~C., McCutchan,
  J.~A., et~al. (2012).
\newblock Low csf concentrations of the nucleotide hiv reverse transcriptase
  inhibitor, tenofovir.
\newblock {\em Journal of acquired immune deficiency syndromes (1999)},
  59(4):376.

\bibitem[Borghetti et~al., 2017]{borghetti2017efficacy}
Borghetti, A., Baldin, G., Capetti, A., SterrantCohen~ino, G., Rusconi, S.,
  Latini, A., Giacometti, A., Madeddu, G., Picarelli, C., De~Marco, R., Cossu,
  M.~V., Lagi, F., Cauda, R., De~Luca, A., Giambenedetto, D., and Group, S.
  O.~S. (2017).
\newblock {Efficacy and tolerability of dolutegravir and two nucleos (t) ide
  reverse transcriptase inhibitors in HIV-1-positive, virologically suppressed
  patients}.
\newblock {\em AIDS}, 31(3):457--459.

\bibitem[Borghetti et~al., 2018]{borghetti2018slc22a2}
Borghetti, A., Calcagno, A., Lombardi, F., Cusato, J., Belmonti, S., D?avolio,
  A., Ciccarelli, N., La~Monica, S., Colafigli, M., Delle~Donne, V., Marco,
  R.~D., Tamburrini, E., Visconti, E., Perri, G.~D., Luca, A.~D., Bonora, S.,
  and Giambenedetto, S.~D. (2018).
\newblock {SLC22A2} variants and dolutegravir levels correlate with psychiatric
  symptoms in persons with {HIV}.
\newblock {\em Journal of Antimicrobial Chemotherapy}.

\bibitem[Brickman et~al., 2017]{brickman2017association}
Brickman, C., Propert, K.~J., Voytek, C., Metzger, D., and Gross, R. (2017).
\newblock Association between depression and condom use differs by sexual
  behavior group in patients with hiv.
\newblock {\em AIDS and Behavior}, 21(6):1676--1683.

\bibitem[Brink et~al., 2010]{brink2010monitoring}
Brink, M.~S., Visscher, C., Arends, S., Zwerver, J., Post, W.~J., and Lemmink,
  K.~A. (2010).
\newblock Monitoring stress and recovery: new insights for the prevention of
  injuries and illnesses in elite youth soccer players.
\newblock {\em British Journal of Sports Medicine}, 44(11):809--815.

\bibitem[Chattopadhyay et~al., 2017]{chattopadhyay2017cognitive}
Chattopadhyay, S., Ball, S., Kargupta, A., Talukdar, P., Roy, K., Talukdar, A.,
  and Guha, P. (2017).
\newblock Cognitive behavioral therapy improves adherence to antiretroviral
  therapy in hiv-infected patients: a prospective randomized controlled trial
  from eastern india.
\newblock {\em HIV \& AIDS Review. International Journal of HIV-Related
  Problems}, 16(2):89--95.

\bibitem[Clubreth et~al., 2016]{clubreth2016associations}
Clubreth, R., Dube, S., and Maggio, D. (2016).
\newblock Associations between major depression, health-risk behaviors, and
  medication adherence among hiv-positive adults receiving medical care in
  georgia.
\newblock {\em Journal of the Georgia Public Health Association}.

\bibitem[Cohen et~al., 2011]{cohen2011randomized}
Cohen, C., Elion, R., Ruane, P., Shamblaw, D., DeJesus, E., Rashbaum, B.,
  Chuck, S.~L., Yale, K., Liu, H.~C., Warren, D.~R., et~al. (2011).
\newblock Randomized, phase 2 evaluation of two single-tablet regimens
  elvitegravir/cobicistat/emtricitabine/tenofovir disoproxil fumarate versus
  efavirenz/emtricitabine/tenofovir disoproxil fumarate for the initial
  treatment of hiv infection.
\newblock {\em Aids}, 25(6):F7--F12.

\bibitem[Cohen et~al., 2017]{cohen2017astrocyte}
Cohen, J., D?Agostino, L., Wilson, J., Tuzer, F., and Torres, C. (2017).
\newblock Astrocyte senescence and metabolic changes in response to hiv
  antiretroviral therapy drugs.
\newblock {\em Frontiers in aging neuroscience}, 9:281.

\bibitem[Cook et~al., 2002]{cook2002effects}
Cook, J.~A., Cohen, M.~H., Burke, J., Grey, D., Anastos, K., Kirstein, L.,
  Palacio, H., Richardson, J., Wilson, T., and Young, M. (2002).
\newblock Effects of depressive symptoms and mental health quality of life on
  use of highly active antiretroviral therapy among hiv-seropositive women.
\newblock {\em JAIDS-HAGERSTOWN MD-}, 30(4):401--409.

\bibitem[Dahl, 2006]{dahl2006model}
Dahl, D.~B. (2006).
\newblock {Model-based clustering for expression data via a Dirichlet process
  mixture model}.
\newblock In Do, K.-A., M{\"u}ller, P., Vannucci, MarinaDo, K.-A., M{\"u}ller,
  P., and Vannucci, M., editors, {\em {Bayesian Inference for Gene Expression
  and Proteomics}}, chapter~10, pages 201--218. Cambridge University Press.

\bibitem[Derogatis et~al., 1974]{derogatis1974hopkins}
Derogatis, L.~R., Lipman, R.~S., Rickels, K., Uhlenhuth, E.~H., and Covi, L.
  (1974).
\newblock {The Hopkins Symptom Checklist (HSCL): A self-report symptom
  inventory}.
\newblock {\em Behavioral science}, 19(1):1--15.

\bibitem[Eilers and Marx, 1996]{eilers1996flexible}
Eilers, P.~H. and Marx, B.~D. (1996).
\newblock Flexible smoothing with b-splines and penalties.
\newblock {\em Statistical {S}cience}, pages 89--102.

\bibitem[Elzi et~al., 2017]{elzi2017adverse}
Elzi, L., Erb, S., Furrer, H., Cavassini, M., Calmy, A., Vernazza, P.,
  G{\"u}nthard, H., Bernasconi, E., and Battegay, M. (2017).
\newblock Adverse events of raltegravir and dolutegravir.
\newblock {\em AIDS (London, England)}, 31(13):1853.

\bibitem[Ferguson, 1974]{ferguson1974prior}
Ferguson, T.~S. (1974).
\newblock Prior distributions on spaces of probability measures.
\newblock {\em The annals of statistics}, 2(4):615--629.

\bibitem[Fried et~al., 2016]{fried2016measuring}
Fried, E.~I., van Borkulo, C.~D., Epskamp, S., Schoevers, R.~A., Tuerlinckx,
  F., and Borsboom, D. (2016).
\newblock Measuring depression over time... or not? lack of unidimensionality
  and longitudinal measurement invariance in four common rating scales of
  depression.
\newblock {\em Psychological Assessment}, 28(11):1354.

\bibitem[Gelman et~al., 2014]{gelman2014understanding}
Gelman, A., Hwang, J., and Vehtari, A. (2014).
\newblock Understanding predictive information criteria for bayesian models.
\newblock {\em Statistics and computing}, 24(6):997--1016.

\bibitem[Harris et~al., 2008]{harris2008exacerbation}
Harris, M., Larsen, G., and Montaner, J.~S. (2008).
\newblock Exacerbation of depression associated with starting raltegravir: a
  report of four cases.
\newblock {\em Aids}, 22(14):1890--1892.

\bibitem[Hoffmann et~al., 2017]{hoffmann2017higher}
Hoffmann, C., Welz, T., Sabranski, M., Kolb, M., Wolf, E., Stellbrink, H.-J.,
  and Wyen, C. (2017).
\newblock Higher rates of neuropsychiatric adverse events leading to
  dolutegravir discontinuation in women and older patients.
\newblock {\em HIV medicine}, 18(1):56--63.

\bibitem[Ickovics et~al., 2001]{ickovics2001mortality}
Ickovics, J.~R., Hamburger, M.~E., Vlahov, D., Schoenbaum, E.~E., Schuman, P.,
  Boland, R.~J., Moore, J., and Group, H. E. R.~S. (2001).
\newblock Mortality, cd4 cell count decline, and depressive symptoms among
  hiv-seropositive women: longitudinal analysis from the hiv epidemiology
  research study.
\newblock {\em Jama}, 285(11):1466--1474.

\bibitem[Ironson et~al., 2017]{ironson2017depression}
Ironson, G., Fitch, C., and Stuetzle, R. (2017).
\newblock Depression and survival in a 17-year longitudinal study of people
  with hiv: Moderating effects of race and education.
\newblock {\em Psychosomatic medicine}, 79(7):749--756.

\bibitem[Jones et~al., 2017]{jones2017training}
Jones, C.~M., Griffiths, P.~C., and Mellalieu, S.~D. (2017).
\newblock Training load and fatigue marker associations with injury and
  illness: a systematic review of longitudinal studies.
\newblock {\em Sports Medicine}, 47(5):943--974.

\bibitem[Kapfhammer, 2006]{kapfhammer2006somatic}
Kapfhammer, H.-P. (2006).
\newblock Somatic symptoms in depression.
\newblock {\em Dialogues in clinical neuroscience}, 8(2):227.

\bibitem[Kroenke et~al., 2001]{kroenke2001phq}
Kroenke, K., Spitzer, R.~L., and Williams, J.~B. (2001).
\newblock {The PHQ-9: validity of a brief depression severity measure}.
\newblock {\em Journal of general internal medicine}, 16(9):606--613.

\bibitem[Lang and Brezger, 2004]{lang2004bayesian}
Lang, S. and Brezger, A. (2004).
\newblock Bayesian p-splines.
\newblock {\em Journal of computational and graphical statistics},
  13(1):183--212.

\bibitem[Lewinsohn et~al., 1997]{lewinsohn1997center}
Lewinsohn, P.~M., Seeley, J.~R., Roberts, R.~E., and Allen, N.~B. (1997).
\newblock {Center for Epidemiologic Studies Depression Scale (CES-D) as a
  screening instrument for depression among community-residing older adults.}
\newblock {\em Psychology and aging}, 12(2):277.

\bibitem[Li et~al., 2019]{li2019bareb}
Li, Y., Bandyopadhyay, D., Xie, F., and Xu, Y. (2019).
\newblock Bareb: A bayesian repulsive biclustering model for periodontal data.
\newblock {\em arXiv preprint arXiv:1902.05680}.

\bibitem[Maki et~al., 2012]{maki2012depressive}
Maki, P.~M., Rubin, L.~H., Cohen, M., Golub, E.~T., Greenblatt, R.~M., Young,
  M., Schwartz, R.~M., Anastos, K., and Cook, J.~A. (2012).
\newblock {Depressive symptoms are increased in the early perimenopausal stage
  in ethnically diverse HIV+ and HIV- women}.
\newblock {\em Menopause (New York, NY)}, 19(11):1215.

\bibitem[Mills et~al., 2016]{mills2016switching}
Mills, A., Arribas, J.~R., Andrade-Villanueva, J., DiPerri, G., Van~Lunzen, J.,
  Koenig, E., Elion, R., Cavassini, M., Madruga, J.~V., Brunetta, J., et~al.
  (2016).
\newblock Switching from tenofovir disoproxil fumarate to tenofovir alafenamide
  in antiretroviral regimens for virologically suppressed adults with hiv-1
  infection: a randomised, active-controlled, multicentre, open-label, phase 3,
  non-inferiority study.
\newblock {\em The Lancet Infectious Diseases}, 16(1):43--52.

\bibitem[Mollan et~al., 2014]{mollan2014association}
Mollan, K.~R., Smurzynski, M., Eron, J.~J., Daar, E.~S., Campbell, T.~B., Sax,
  P.~E., Gulick, R.~M., Na, L., O'Keefe, L., Robertson, K.~R., et~al. (2014).
\newblock Association between efavirenz as initial therapy for hiv-1 infection
  and increased risk for suicidal ideation or attempted or completed suicide:
  an analysis of trial data.
\newblock {\em Annals of internal medicine}, 161(1):1--10.

\bibitem[Mollan et~al., 2017]{mollan2017race}
Mollan, K.~R., Tierney, C., Hellwege, J.~N., Eron, J.~J., Hudgens, M.~G.,
  Gulick, R.~M., Haubrich, R., Sax, P.~E., Campbell, T.~B., Daar, E.~S.,
  Robertson, K.~R., Ventura, D., Ma, Q., Edwards, D. R.~V., Haas, D.~W., and
  the AIDS Clinical Trials~Group (2017).
\newblock Race/ethnicity and the pharmacogenetics of reported suicidality with
  efavirenz among clinical trials participants.
\newblock {\em The Journal of infectious diseases}, 216(5):554--564.

\bibitem[Moore et~al., 1999]{moore1999severe}
Moore, J., Schuman, P., Schoenbaum, E., Boland, B., Solomon, L., and Smith, D.
  (1999).
\newblock Severe adverse life events and depressive symptoms among women with,
  or at risk for, hiv infection in four cities in the united states of america.
\newblock {\em Aids}, 13(17):2459--2468.

\bibitem[M{\"u}ller and Quintana, 2004]{muller2004nonparametric}
M{\"u}ller, P. and Quintana, F.~A. (2004).
\newblock Nonparametric bayesian data analysis.
\newblock {\em Statistical science}, pages 95--110.

\bibitem[Nanni et~al., 2015]{nanni2015depression}
Nanni, M.~G., Caruso, R., Mitchell, A.~J., Meggiolaro, E., and Grassi, L.
  (2015).
\newblock {Depression in HIV infected patients: a review}.
\newblock {\em Current psychiatry reports}, 17(1):530.

\bibitem[Revuelta-Herrero et~al., 2018]{revuelta2018effectiveness}
Revuelta-Herrero, J.~L., Chamorro-de Vega, E., Rodr{\'\i}guez-Gonz{\'a}lez,
  C.~G., Alonso, R., Herranz-Alonso, A., and Sanjurjo-S{\'a}ez, M. (2018).
\newblock Effectiveness, safety, and costs of a treatment switch to
  dolutegravir plus rilpivirine dual therapy in treatment-experienced hiv
  patients.
\newblock {\em Annals of Pharmacotherapy}, 52(1):11--18.

\bibitem[Rubin et~al., 2011]{rubin2011perinatal}
Rubin, L.~H., Cook, J.~A., Grey, D.~D., Weber, K., Wells, C., Golub, E.~T.,
  Wright, R.~L., Schwartz, R.~M., Goparaju, L., Cohan, D., et~al. (2011).
\newblock {Perinatal depressive symptoms in HIV-infected versus HIV-uninfected
  women: a prospective study from preconception to postpartum}.
\newblock {\em Journal of Women's Health}, 20(9):1287--1295.

\bibitem[Shah et~al., 2016]{shah2016neurotoxicity}
Shah, A., Gangwani, M.~R., Chaudhari, N.~S., Glazyrin, A., Bhat, H.~K., and
  Kumar, A. (2016).
\newblock Neurotoxicity in the post-haart era: caution for the antiretroviral
  therapeutics.
\newblock {\em Neurotoxicity research}, 30(4):677--697.

\bibitem[Squires et~al., 2003]{squires2003tenofovir}
Squires, K., Pozniak, A.~L., Pierone, G., Steinhart, C.~R., Berger, D., Bellos,
  N.~C., Becker, S.~L., Wulfsohn, M., Miller, M.~D., Toole, J.~J., et~al.
  (2003).
\newblock Tenofovir disoproxil fumarate in nucleoside-resistant hiv-1
  infection: a randomized trial.
\newblock {\em Annals of internal medicine}, 139(5\_Part\_1):313--320.

\bibitem[Taibi, 2013]{taibi2013sleep}
Taibi, D.~M. (2013).
\newblock Sleep disturbances in persons living with hiv.
\newblock {\em Journal of the Association of Nurses in AIDS Care},
  24(1):S72--S85.

\bibitem[Taniguchi et~al., 2014]{taniguchi2014depression}
Taniguchi, T., Shacham, E., {\"O}nen, N.~F., Grubb, J.~R., and Overton, E.~T.
  (2014).
\newblock Depression severity is associated with increased risk behaviors and
  decreased cd4 cell counts.
\newblock {\em AIDS care}, 26(8):1004--1012.

\bibitem[Underwood et~al., 2015]{underwood2015could}
Underwood, J., Robertson, K.~R., and Winston, A. (2015).
\newblock Could antiretroviral neurotoxicity play a role in the pathogenesis of
  cognitive impairment in treated hiv disease?
\newblock {\em Aids}, 29(3):253--261.

\bibitem[Watanabe, 2010]{watanabe2010asymptotic}
Watanabe, S. (2010).
\newblock Asymptotic equivalence of bayes cross validation and widely
  applicable information criterion in singular learning theory.
\newblock {\em Journal of Machine Learning Research}, 11(Dec):3571--3594.

\bibitem[Williams et~al., 2020]{williams2020associations}
Williams, D.~W., Li, Y., Dastgheyb, R., Fitzgerald, K.~C., Maki, P.~M., Spence,
  A.~B., Gustafson, D.~R., Milam, J., Sharma, A., Adimora, A.~A., et~al.
  (2020).
\newblock Associations between antiretroviral drugs on depressive
  symptomatology in homogenous subgroups of women with hiv.
\newblock {\em Journal of Neuroimmune Pharmacology}, pages 1--14.

\bibitem[Xu et~al., 2016]{xu2016non}
Xu, Y., Xu, Y., and Saria, S. (2016).
\newblock A non-parametric {B}ayesian approach for estimating
  treatment-response curves from sparse time series.
\newblock In {\em Proceedings of the 1st Machine Learning for Healthcare
  Conference}, pages 282--300.

\bibitem[Zash et~al., 2018]{zash2018neural}
Zash, R., Makhema, J., and Shapiro, R.~L. (2018).
\newblock Neural-tube defects with dolutegravir treatment from the time of
  conception.
\newblock {\em New England Journal of Medicine}, 379(10):979--981.

\end{thebibliography}

\end{document}